\documentclass[journal]{IEEEtran}

\usepackage{graphicx} 

\usepackage{amssymb} 

\usepackage{lineno} 

\usepackage[hyphens]{url} 

\usepackage{enumitem} 

\usepackage{subfig} 

\usepackage[export]{adjustbox} 

\usepackage{hhline} 

\usepackage{multirow} 
\usepackage{booktabs} 

\usepackage{amsmath,amssymb,amsfonts} 

\usepackage{bmpsize}

\usepackage{bm} 

\usepackage{bbding} 

\usepackage{makecell} 

\usepackage{color,soul} 

\ifCLASSINFOpdf
\else
\fi

\usepackage{dblfloatfix}

\usepackage
{soul} 

\begin{document}

\title{D-Score: An Expert-Based Method for Assessing the Detectability of IoT-Related Cyber-Attacks}

\author{Yair~Meidan,
        Daniel~Benatar,
        Ron~Bitton, 
        Dan Avraham,
        and~Asaf~Shabtai
\thanks{All of the authors are with the Department
of Software and Information Systems Engineering, Ben-Gurion University of the Negev, Beer-Sheva,
Israel, e-mail: \{yairme, benatar, ronbit, danavra\}@post.bgu.ac.il, shabtaia@bgu.ac.il.}
\thanks{This project has received funding from the European Union’s Horizon 2020 research and innovation programme under grant agreement No 830927.}
\thanks{Manuscript received December 3, 2021; revised August 5, 2022.}}

\markboth{Computers \& Security,~Vol.~VV, No.~NN, MM~YYYY}%
{Author \MakeLowercase{\textit{et al.}}: D-Score}


\maketitle

\begin{abstract}
IoT devices are known to be vulnerable to various cyber-attacks, such as data exfiltration and the execution of flooding attacks as part of a DDoS attack. When it comes to detecting such attacks using network traffic analysis, it has been shown that some attack scenarios are not always equally easy to detect if they involve different IoT models. That is, when targeted at some IoT models, a given attack can be detected rather accurately, while when targeted at others the same attack may result in too many false alarms. In this research, we attempt to explain this variability of IoT attack detectability and devise a risk assessment method capable of addressing a key question: \emph{how easy is it for an anomaly-based network intrusion detection system to detect a given cyber-attack involving a specific IoT model?} In the process of addressing this question we \emph{(a)} investigate the predictability of IoT network traffic, \emph{(b)} present a novel taxonomy for IoT attack detection which also encapsulates traffic predictability aspects, \emph{(c)} propose an expert-based attack detectability estimation method which uses this taxonomy to derive a detectability score (termed \emph{`D-Score'}) for a given combination of IoT model and attack scenario, and \emph{(d)} empirically evaluate our method while comparing it with a data-driven method.
\end{abstract}


\IEEEpeerreviewmaketitle

\section{Introduction}\label{sec:introduction}

The \emph{Internet of Things} (IoT) is a rapidly evolving trend in wireless communication~\cite{da2014internet}, where various objects are connected to the Internet and can cooperate with one another to reach common goals.
In home or enterprise environments, such objects may include \emph{smart} webcams, light bulbs, and motion detectors.
Although designed to improve various aspects of modern life, IoT devices have also become known as easy targets for various cyber-attacks~\cite{hallman2017ioddos, yang2017survey} which continuously increase~\cite{kaspersky2019, symantec2019} along with the worldwide proliferation of the IoT~\cite{gartner2019, idc2019}.

\begin{figure*}[t]
\begin{minipage}{.58\linewidth}
\centering
\subfloat[]{\label{subfig:method_overview}\includegraphics[height=0.185\textheight, trim={0cm 6.6cm 0cm 0cm},clip, right]
{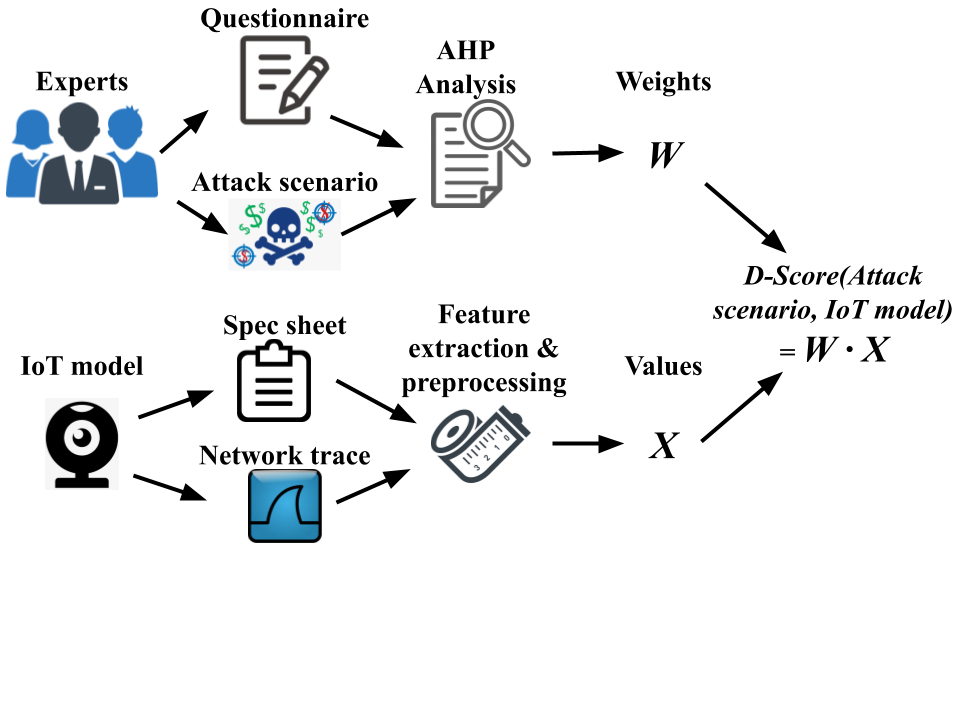}}
\end{minipage}
\hspace{\fill}
\begin{minipage}{.38\linewidth}
\centering
\subfloat[]{\label{subfig:label_example}\includegraphics[height=0.185\textheight
, trim={-2cm -2cm 0.65cm -2cm},clip
]
{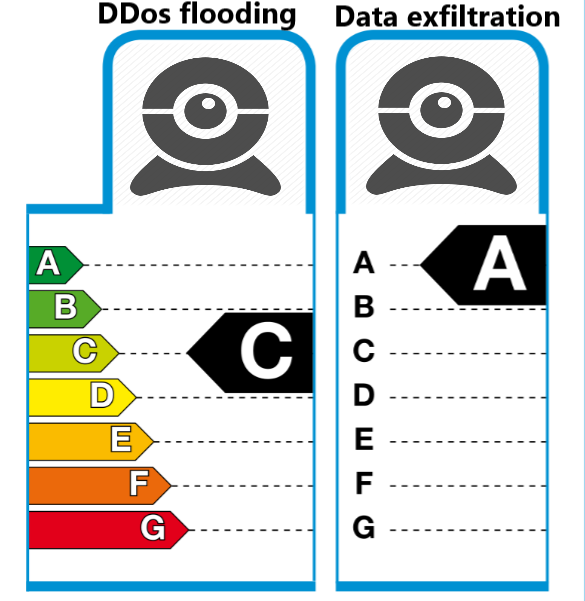}}
\end{minipage}
\caption{Overview of \emph{(a)} the proposed method for calculating D-Score and \emph{(b)} the derived D-Score label, inspired by~\cite{wikityre}.}
\label{fig:method_and_label}
\end{figure*}

A fundamental problem in IoT security is that usually,~\cite{kaspersky2019} attacks on IoT devices are stealthy. On top of that, anomaly-based network intrusion detection systems (AIDSs)~\cite{garcia2009anomaly} might under-perform if the normal traffic behavior is complex and thus challenging to characterize. That is, when a relatively complex IoT device is compromised by, e.g., an IoT botnet, the malicious traffic might be camouflaged by the already-irregular benign traffic, such that the malicious traffic would not be deemed anomalous by the AIDS. This could happen due to high inherent variability in the device's \emph{dynamic} features (e.g., diversification of source and destination IP addresses, unsteadiness of packet sizes and inter-arrival times, etc.), which are possibly influenced by the device's \emph{static} features (e.g., quantity and diversity of sensors and actuators, memory size and CPU speed).

Overall, due to either the attack's stealthiness or the normal traffic's irregularity, users may not notice that their devices are being exploited, so these devices might remain connected to networks and pose an ongoing threat. Therefore, home users and an enterprise network security administrators are advised to be cautious when deploying IoT devices which, once exploited, the attacks they might execute are hard to detect.

In many studies (surveyed in~\cite{yang2017survey, zhao2013survey,
alaba2017internet}), machine learning methods have been proposed as means for AIDSs to monitor ongoing network traffic data and differentiate benign IoT activity and malicious events. The empirical results of such research papers typically demonstrate promising \emph{overall} classification performance. However, as noted in~\cite{meidan2018nbaiot, Bahsi2018Dimensionality, nomm2018unsupervised}, the classification performance of an AIDS may vary among different IoT models even when they are compromised in the same way. This phenomenon of variability of IoT attack detection performance leads to the following two groups of interesting questions which we attempt to answer throughout this research:
\begin{enumerate}[leftmargin=*]
    \item \textbf{Network traffic predictability:} Is there a difference between the regularity/predictability of IoT network traffic and non-IoT traffic? How variable is the traffic predictability among various IoT models? Can we correlate between IoT model complexity (i.e., static features) and traffic predictability (dynamic features)?
    
    \item \textbf{IoT attack detectability:} For IoT devices, can we explain why a given attack is relatively easy to detect on one IoT model, yet difficult to detect on another by the same AIDS? Can the ability to detect a given attack scenario involving a given IoT model be quantified in advance, and standardized into an IoT security index or label?
\end{enumerate}

Past related studies have already attempted at developing standard IoT security indices~\cite{blythe2018consumer} and designing related labels~\cite{kelley2009nutrition}. However, none of them measures attacks detectability. In this study, inspired by the notion of \emph{IoT traffic predictability}~\cite{meidan2018nbaiot, doshi2018machine, Bertino2017BotnetsSecurity, bezerra2019iotds, cvitic2020definition}, we address the above questions and coin a new complementary term, namely, the \emph{IoT Attack Detectability Score} (abbreviated as \emph{``D-Score''}).
The D-Score is an assessment as to the ease of detecting IoT-related cyber-attacks by an AIDS.
It is calculated for combinations of an attack scenario ($A$) and IoT model ($M$), as outlined in Eq.~\eqref{eq:d_score_structure}:
\begin{equation}
D-Score: A \times M \Rightarrow \mathbb{R} \label{eq:d_score_structure}
\end{equation}

To calculate the D-Score, a data-driven approach can be considered, where machine learning-based anomaly detectors are first trained and then applied to various IoT models in order to capture the relationship between the AIDS's performance and the IoT model's static and/or dynamic features. However, in practice~\cite{singla2019overcoming}, this approach requires large amounts of labeled training data that are often expensive~\cite{alaei2017incremental} and time-consuming to acquire (see Subsection \ref{subsec:expert_vs_data}). As an alternative, in this paper we propose an expert-based method (outlined in Fig.~\ref{subfig:method_overview}, elaborated in Section~\ref{sec:proposed_method}). Generally, for an attack scenario $A$, cyber-security experts fill in a questionnaire which is constructed and analyzed in accordance with the analytical hierarchical process (AHP)~\cite{saaty1988analytic}, to obtain feature weights $W$. Then, for an IoT model $M$, static and dynamic feature values $X$ are obtained from the spec sheet and network traces, respectively. Eventually, the D-Score is calculated as a weighted sum ($W \cdot X$). The resultant scalar ranges from zero (``impossible to detect an attack'') to one (``easy to detect''), and it can be translated into a detectability \emph{label} (see Fig.~\ref{subfig:label_example}). Once D-Score labels are associated with IoT devices, domestic and enterprise customers will be more able to make informed decisions.

We summarize our contributions as follows:
\begin{enumerate}[leftmargin=*]
    \item Numerous past studies have relied on the (reasonable) hypothesis that the network traffic of IoT devices is significantly more steady and predictable than the traffic of non-IoT devices (computers, smartphones, etc.). Compared to them, we are the first to quantitatively investigate this hypothesis, using two datasets. In addition, we examine the variability of traffic predictability among disparate IoT models, correlate this variability with static features of IoT models, and leverage the variability in traffic predictability to quantify in advance IoT attack detectability.
    \item We are the first (to the best of our knowledge) to explore the quantification of IoT attack detectability in an explicit and systematic manner. Note that in contrast to past studies which addressed the ease of \emph{compromising} an IoT device~\cite{anand2020iotVulnerability} and \emph{executing} a cyber-attack, no past study attempted to quantify the ease of \emph{detecting} such attacks if and when they are executed.
    \item To quantify IoT attack detectability in advance, we propose a novel expert-based method, which could spare some technically-challenging and time-consuming actions, mostly attack implementation and precise ground-truth labeling. As a basis for our method, we present a novel designated taxonomy, and for each feature in the taxonomy we also propose a means of calculating its value for a given IoT model.
    \item To facilitate the collection of responses from cyber-security experts, we developed an online questionnaire~\footnote{https://dscore.limequery.com/915153}. This questionnaire incorporates two novel elements, namely \emph{(a)} preliminary filtering of categories and sub-categories, and \emph{(b)} dynamic selection of the attack scenario to be addressed by the respondent.
    The former minimizes the number of comparisons and thus reduces the burden on respondents; the latter facilitates the addition of attack scenarios in future research, without the need to implement and maintain multiple versions of the questionnaire.
    \item The questionnaire, which addresses 4 common IoT-related attack scenarios, was completed by 40 cyber-security experts from various organizations (both from academia and industry) in the European Union, the Middle East and the Far East. 
    We share~\footnote{http://doi.org/10.5281/zenodo.4018614} the set of (anonymous) responses with the research community so it can be utilized in future research.
\end{enumerate}

\section{Use Cases, Scope and Assumptions}
\label{sec:scope_assumptions_and_use_cases}

\subsection{Use Cases}\label{subsec:Use_Cases}

\begin{enumerate}[leftmargin=*]
    \item Looking backward, a D-Score can provide guidance for specialists of why the same attack was detected with varying levels of accuracy for diverse IoT models. In academic settings, this can help explaining the variability in attack detection performance among experimented IoT devices~\cite{meidan2018nbaiot, Bahsi2018Dimensionality, nomm2018unsupervised}. In enterprise settings, the D-Score can assist in analysing the functioning of an AIDS in IoT-related cyber-incidents. 
    \item Looking forward, a D-Score can be leveraged to support procurement decisions of home users and enterprise network administrators. For instance, when contemplating which \emph{smart} security camera model to deploy, a model with a higher D-Score is expected to demonstrate better detection performance, thus safer and more advisable. The D-Score can also supplement organizational processes of risk assessment and risk mitigation. Further details are provided in Subsection~\ref{subsec:risk_assessment_iot} and Section~\ref{sec:proposed_method}, respectively.
    \end{enumerate}

\subsection{Scope}\label{subsec:scope}

In this research we propose a taxonomy and a method to assess in advance the attack detectability (rather than device exploitability) of IoT models. To this end, similarly to~\cite{meidan2020deNAT}, we define an \emph{IoT model} as a combination of four elements: Type/manufacturer/model number/firmware version. For instance, Webcam/Provision/PT-838/v.1.3.3~\cite{PT838E} is an IoT model we use in this research for evaluation. We focus on the IoT model granularity rather than, e.g., the device's manufacturer or type, because in many cases~\cite{meidan2020deNAT, miettinen2017iot, antonakakis2017understanding} IoT malware rely on exploiting vulnerabilities that are strongly associated with specific firmware versions of specific models made by specific manufacturers. From among the various application domains in which the IoT is proliferating~\cite{perwej2019internet}, we concentrate on the sectors of \emph{smart} homes and enterprises, and relate mostly to high-end commercial IoT devices such as web-enabled cameras, TVs, light bulbs, sockets, baby monitors, thermostats, door bells, etc. Those devices typically connect to the Internet via Wi-Fi and transfer data using the TCP/IP stack. Our taxonomy (see Section~\ref{sec:taxonomy}) is tailored to reflect those typical characteristics.

\subsection{Assumptions}\label{subsec:assumptions}

Considering the scope defined above, we make several assumptions in this research. First, we assume that there is concern that a stealthy IoT-related attack might occur and go undetected by an AIDS. More specifically, we address attacks that can potentially be detected by an AIDS (though with various detectability levels), such as data exfiltration~\cite{d2016data, irion2017weeping}, botnet gathering~\cite{Bertino2017BotnetsSecurity}, communication with a command and control (C\&C) server~\cite{hallman2017ioddos}, and execution of flooding attacks as part of a distributed denial of service (DDoS) attack~\cite{hallman2017ioddos}. We also assume that the IoT device is connected to the Internet directly (i.e., not via a communal gateway) and that its network traffic data is being monitored by either enterprise-level~\cite{ullah2018protection} or domestic~\cite{allot,avast} intrusion detection systems.

\subsection{Threat Model}\label{subsec:threat_model}
In this research we assume a system in which an IoT device is connected to a local area network (LAN) and the Internet. We also assume that the LAN is monitored by an AIDS in order to identify a variety of cyber-attacks. In this research we consider two threat models and evaluate them quantitatively (see Section~\ref{sec:evaluation_method_and_validation}. In the first threat model, we consider an infected IoT device that is fully controlled by the attacker. In this threat model, we assume that the attacker can receive/transmit network packets from/to the LAN and the Internet. By utilizing these capabilities, the attacker can execute the following attacks (summarized in Table~\ref{tab:attack_scenarios}): \emph{(a) C\&C communication} where the infected IoT device tries to communicate with a C\&C server controlled by the attacker and located on the Internet. To avoid detection of C\&C communication, adversaries often mimic normal traffic. \emph{(b) DDoS flooding attacks} where the infected IoT devices are used to perform network denial-of-service attacks. This is done by exhausting the network bandwidth of the target service (such as specific websites, email services, the DNS, and web-based applications). \emph{(c) Data exfiltration attacks} where the attacker can scan the target network and send reconnaissance data to a server controlled by the attacker. To do so, adversaries often utilize compression and encryption, and they can also set size limits on the data transmission to avoid detection. In the second threat model, we consider a legitimate IoT device that is deployed on the target network. We further assume that the IoT device is reachable from the Internet. In this threat model, we consider a \emph{bot scanning} scenario where an adversary scans the IoT device for recruitment as part of botnet propagation. Note that the threat models we assume and evaluate can easily be extended in the future, without the need to adjust the taxonomy presented in this paper (see Section~\ref{sec:taxonomy}).

\section{\label{sec:background_and_related_work}Background and Related Work}

\subsection{\label{subsec:iot_vs_non_iot_security}IoT vs. Non-IoT Security}

Compared with the cyber-security aspects of other, non-IoT, systems, IoT security is known for its large attack surface. This attack surface exists on billions of devices, many of which are unfortunately known to have increased vulnerability~\cite{anand2020iotVulnerability}. Broad heterogeneity is one of the common attributes used for the characterization of IoT devices, making it difficult for device manufacturers to institute a single standard of security, unlike traditional IT systems, which are less heterogeneous. Moreover, compared to traditional IT systems, an IoT deployment may involve numerous devices and countless lines of code created by a much larger pool of developers and have many types of hardware and operating systems. Also, traditional IT devices such as local servers and personal computers are typically located in a closed environment with access control measures, whereas IoT devices are often located in open and unattended environments, leaving  an opening for intruders to intentionally gain physical access to them~\cite{varshney2019architectural}. IoT devices are often resource-constrained in terms of processing power and memory, such that they might lack robust security protocols, and sufficient computational power for encryption~\cite{jing2014security}. Due to these hardware limitations, many smart devices are not capable of protecting themselves with host-based security solutions such as endpoint protection platforms, anti-malware, and endpoint detection and response. These solutions, while well suited for traditional IT systems, are too computationally intensive for the limited power, battery, and memory capacity of IoT devices. Unlike servers and PCs that constantly receive security updates and patches, IoT devices are updated infrequently and fail to address the risk of zero-day vulnerabilities. Moreover, in many cases, end users are not capable of adding security measures to the built-in operational system. Lastly, IoT devices depend on third-party libraries and components that act as a `black box,' making them difficult to patch, control, and scan for vulnerabilities.

\subsection{\label{subsec:iot_security}Recent Security Research in the IoT}

IoT security is a broad area of research, in which a variety of subjects are being studied. These subjects include various IoT attack scenarios, defense mechanisms, risk mitigation approaches, and risk assessment methods. In multiple recent studies on IoT security, several attacks are repeatedly mentioned as open-ended key challenges. Among these these attacks are DDoS attacks~\cite{roopak2020multi} (addressed in our research as well) which are considered a convenient way for an attacker to abuse a system due to the large number of IoT devices that are continuously connected to the Internet~\cite{shaukat2021review}. In a recent review paper on IoT security challenges~\cite{sutar2022extensive}, DDoS attacks and data sniffing are mentioned as major threats to IoT systems, while in~\cite{papalkar2022analysis}, the authors proposed a deep learning-based methodology for DDoS attack detection and zero-day attacks in IoT networks. Another IoT-related attack that has been addressed in several recent studies (including our research) is data exfiltration. This type of attack is commonly considered a big security challenge~\cite{leevy2022iot,uroz2022characterization}. Another prominent security issue in IoT systems is bot detection, a challenging task that has received a lot of attention from researchers, including~\cite{jayalaxmi2022debot}, where deep learning techniques are used for bot detection in IoT network traffic.

\subsection{\label{subsec:risk_assessment_iot}Risk Assessment in the IoT}

Along with the IoT expansion, the demand for appropriate risk assessment methods has increased. However, due to the heterogeneity, scalability and connectivity of IoT devices, IoT risk assessment presents significant challenges. Over the years, various methods have been developed as a means of evaluating the security risks associated with the IoT. Among those methods, IoT attack graphs~\cite{shivraj2017graph, mathov2019challenges,
agmon2019deployment} provide a comprehensive view of attacks, and 
can assist in identifying multi-hop high-risk attack paths in a large-scale environment. A different objective was addressed in~\cite{siboni2019weighted}, where the authors aimed at quantifying the ease of IoT-related attack execution. In more recent research studies~\cite{siboni2020ranking, shaghaghi2021iot}, the authors attempted at developing a method for IoT device risk assessment, i.e., for quantifying in advance the influence of an IoT device on the security level of the environment within which it operates. Unlike these and other studies, our goal is not to identify IoT attack paths, determine the ease of \emph{compromising} an IoT device or estimate an IoT device's influence on the security level of its environment. In contrast, our goal is to quantify in advance the ease of \emph{detecting} an IoT attack, assuming that an attack of this kind is likely to occur.

\subsection{Existing Taxonomies in IoT Security}\label{subsec:existing_taxonomies}

For the quantification of IoT attack detectability, the methodology we propose relies on two elements: \emph{(a)} a novel taxonomy, and \emph{(b)} an expert-based mechanism that assigns a weight to each feature in the taxonomy.
Regarding the first element, several taxonomies have already been proposed, none of which sufficiently supports our goal. For example, none of the existing taxonomies addresses user interaction factors, which are strongly correlated~\cite{apthorpe2019keeping, subahi2019detecting} with traffic generation. 

Most existing IoT security-related taxonomies can be assigned to one of the following two categories:  \emph{(a)} IoT architecture taxonomies~\cite{lu2018internet}, or \emph{(b)} IoT attack mechanism taxonomies~\cite{Hossain2017SecuringProblems, nawir2016iot}. 

\textbf{IoT architecture taxonomies} are challenging to define, mainly due to the heterogeneous nature of IoT devices and the lack of standardization in this domain~\cite{architecture2016AlQaseemi}. 
In addition, IoT architectures may need to be more adaptive than traditional computing architectures, in order to cope with the rapid advancement in this field as well as the typical real-time interaction of a given IoT device within its environment. 
In~\cite{lu2018internet}, a four-layered IoT architecture was suggested, and each layer was associated with various attack types. The authors of~\cite{lu2018internet} showed how previously proposed IoT architectures rely on the OSI model~\cite{iso1989} using varying granularity levels. These architectures are relatively general (rather than attack-oriented), as they mostly discuss the structure of the layers, their functionality, and their main components.

\begin{figure*}[ht]
\begin{minipage}{.45\linewidth}
\centering
\subfloat[]{\label{subfig:IN_PKTS_IoT_vs_Non_IoT}\includegraphics[height=0.22\textheight, trim={0cm 0cm 0cm 0cm},clip]
{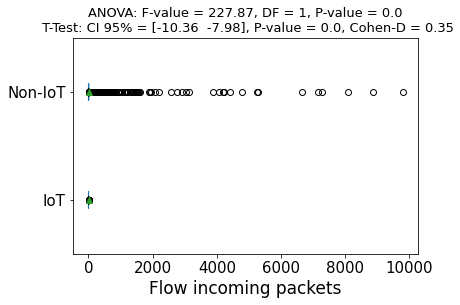}}
\end{minipage}
\hspace{\fill}
\begin{minipage}{.45\linewidth}
\centering
\subfloat[]{\label{subfig:N_flows_per_hour_IoT_vs_Non_IoT}\includegraphics[height=0.22\textheight, trim={0cm 0cm 0cm 0cm},clip]
{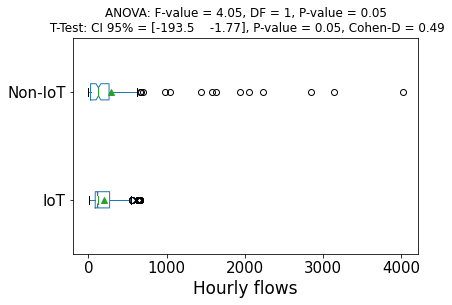}}
\end{minipage}
\vskip\baselineskip
\begin{minipage}{.45\linewidth}
\centering
\subfloat[]{\label{subfig:N_distinct_IPV4_DST_ADDR_per_hour_IoT_vs_Non_IoT}\includegraphics[height=0.22\textheight, trim={0cm 0cm 0cm 0cm},clip]
{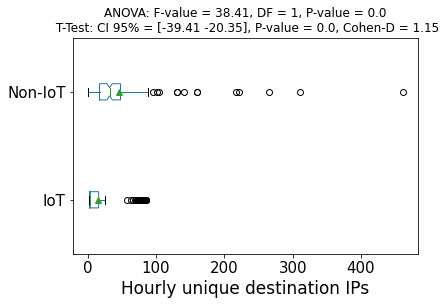}}
\end{minipage}
\hspace{\fill}
\begin{minipage}{.45\linewidth}
\centering
\subfloat[]{\label{subfig:N_distinct_L4_DST_PORT_per_hour_IoT_vs_Non_IoT}\includegraphics[height=0.22\textheight, trim={0cm 0cm 0cm 0cm},clip]
{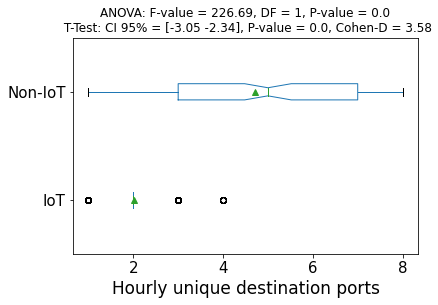}}
\end{minipage}
\caption{Comparison of network traffic predictability between IoT and non-IoT devices.}
\label{fig:iot_vs_non_iot}
\end{figure*}

\textbf{IoT attack mechanism taxonomies} are also used to characterize IoT devices, however with an emphasis on related attacks.
In~\cite{inproceedingsRizvi2018}, the authors focused on applying traditional Internet standards to smart devices, in order to simplify the integration of cyber-attacks in the IoT context. 
In contrast,~\cite{Hossain2017SecuringProblems} and~\cite{nawir2016iot} studied the limitations of traditional security when applied to smart devices, and suggested IoT-specific taxonomies of security attacks. The authors of~\cite{Hossain2017SecuringProblems} presented six elements to consider when characterizing IoT attack mechanisms: device properties, adversary location, attack strategy, access level, information damage level, and host compromise. 
In~\cite{nawir2016iot} two additional elements were suggested: the attack protocol and the communication stack protocol.

Of the above two categories, our study is more closely related to the latter, as we present a security-oriented IoT taxonomy. In comparison to existing taxonomies, we add behavioral and IoT-specific features in a measurable manner, thus facilitate the quantification of IoT attack detectability. In contrast to existing taxonomies, ours also proposes a hierarchical structure that serves to characterize a given IoT model from a broad range of viewpoints, i.e., hardware, software, user interaction (not addressed before in the literature), and networking. Moreover, we propose an expert-based means of weighing the features in this novel taxonomy for various attack scenarios.

\begin{figure*}[t]
\begin{minipage}{.45\linewidth}
\centering
\subfloat[]{\label{subfig:IN_PKTS_within_IoT_only}\includegraphics[height=0.17\textheight, trim={0cm 0cm 0cm 0cm},clip]
{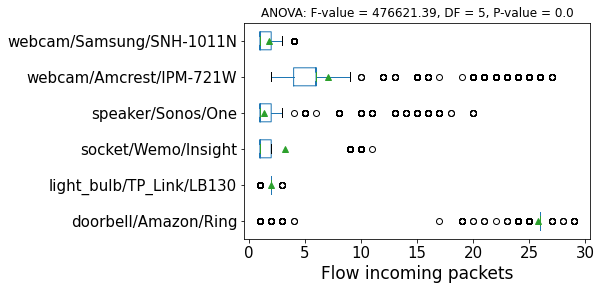}}
\end{minipage}
\hspace{\fill}
\begin{minipage}{.45\linewidth}
\centering
\subfloat[]{\label{subfig:N_flows_per_hour_within_IoT_only}\includegraphics[height=0.17\textheight, trim={0cm 0cm 0cm 0cm},clip]
{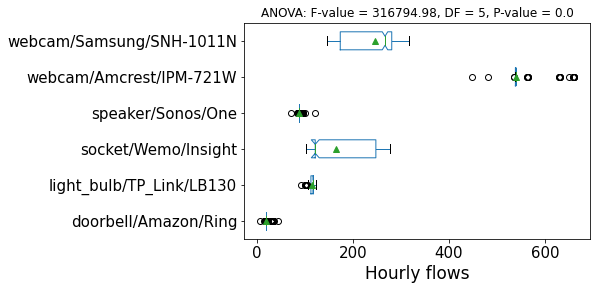}}
\end{minipage}
\vskip\baselineskip
\begin{minipage}{.45\linewidth}
\centering
\subfloat[]{\label{subfig:N_distinct_IPV4_DST_ADDR_per_hour_within_IoT_only}\includegraphics[height=0.17\textheight, trim={0cm 0cm 0cm 0cm},clip]
{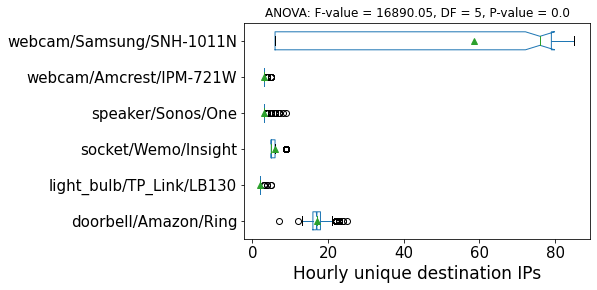}}
\end{minipage}
\hspace{\fill}
\begin{minipage}{.45\linewidth}
\centering
\subfloat[]{\label{subfig:N_distinct_L4_DST_PORT_per_hour_within_IoT_only}\includegraphics[height=0.17\textheight, trim={0cm 0cm 0cm 0cm},clip]
{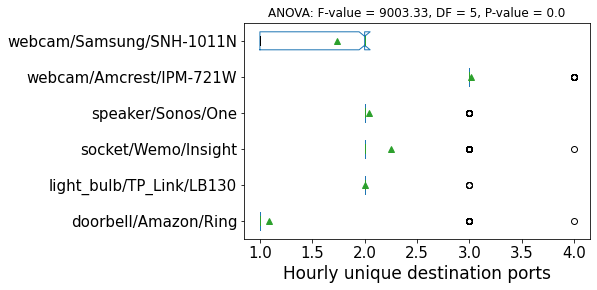}}
\end{minipage}
\caption{Comparison of network traffic predictability among IoT models.}
\label{fig:traffic_predictability_within_iot}
\end{figure*}

\section{Predictability of IoT Network Traffic}\label{sec:iot_traffic_predictability}

\begin{table*}[!b]
\centering
\caption{IoT and non-IoT models used to quantitatively evaluate the traffic predictability, based on the IoT-deNAT dataset~\cite{meidan_yair_2020_3924770}}
\label{tab:model_stats}
\resizebox{\textwidth}{!}{%
\begin{tabular}{l|l|r|r|r|r|c}
\hhline{=======}
\multicolumn{1}{c|}{\textit{\textbf{Category}}} &
  \multicolumn{1}{c|}{\textit{\textbf{Model}}} &
  \multicolumn{1}{c|}{\textit{\textbf{\begin{tabular}[c]{@{}c@{}}Number of\\ outbound flows\end{tabular}}}} &
  \multicolumn{1}{c|}{\textit{\textbf{\begin{tabular}[c]{@{}c@{}}Number of\\ sensors\end{tabular}}}} &
  \multicolumn{1}{c|}{\textit{\textbf{\begin{tabular}[c]{@{}c@{}}Number of\\ actuators\end{tabular}}}} &
  \multicolumn{1}{c|}{\textit{\textbf{\begin{tabular}[c]{@{}c@{}}CPU speed\\ (MHz)\end{tabular}}}} &
  \textit{\textbf{\begin{tabular}[c]{@{}c@{}}Supports\\ adding apps\end{tabular}}} \\ \hhline{=======}
\multirow{6}{*}{IoT}     & webcam/Amcrest/IPM-721W            & 479,492 & 5 & 1 & 1,000 & True                 \\ \cline{2-7} 
                         & webcam/Samsung/SNH-1011N           & 217,899 & 5 & 1 & 216   & False                \\ \cline{2-7} 
                         & light\_bulb/TP\_Link/LB130         & 102,767 & 0 & 0 & 200   & True                 \\ \cline{2-7} 
                         & socket/Wemo/Insight                & 96,426  & 1 & 1 & 360   & True                 \\ \cline{2-7} 
                         & speaker/Sonos/One                  & 78,480  & 0 & 1 & 400   & True                 \\ \cline{2-7} 
                         & doorbell/Amazon/Ring               & 17,179  & 4 & 2 & 80    & True                 \\ \hhline{=======}
\multirow{3}{*}{Non-IoT} & laptop/Dell/Latitude\_E6430        & 23,254  &   &   &       & \multicolumn{1}{r}{} \\ \cline{2-7} 
                         & smartphone/Samsung/Galaxy\_Note\_5 & 15,001  &   &   &       & \multicolumn{1}{r}{} \\ \cline{2-7} 
                         & laptop/Dell/Latitude\_7400         & 4,864   &   &   &       & \multicolumn{1}{r}{} \\ \hhline{=======}
\end{tabular}%
}
\end{table*}

The notion of \emph{IoT traffic predictability} was mentioned several times in the literature~\cite{meidan2018nbaiot, doshi2018machine, Bertino2017BotnetsSecurity, bezerra2019iotds, cvitic2020definition}, however mostly as an (educated) assumption, i.e., without validating it. Throughout this section we quantitatively investigate this assumption and review related literature. For the quantitative investigation we utilize the IoT-deNAT dataset~\cite{meidan_yair_2020_3924770}, which contains (in the form of NetFlows) the outbound traffic generated by commercial IoT models as well as non-IoT devices (see Table~\ref{tab:model_stats}). Note that for simplicity (and also since the firmware of the IoT devices did not change during the data collection period~\cite{meidan2020deNAT}), throughout this paper we omit the \emph{firmware version} component from the IoT model names. Further details about this dataset, which was captured during a substantial period of 37 days, are provided in~\cite{meidan2020deNAT}.

In our experiments we explored the IoT network traffic predictability from three different perspectives: \emph{(a}) Comparison between the traffic predictability of IoT vs. non-IoT devices as two distinct groups, \emph{(b}) comparison of traffic predictability among different models of IoT devices, and \emph{(c}) association of IoT model complexity with traffic predictability, later to be translated into attack detectability. Following are the results of those experiments, in terms of four key performance indicators (KPIs) of network traffic:
\begin{itemize}[leftmargin=*]
    \item \emph{Flow incoming packets}: Number of packets in an inbound flow
    \item \emph{Hourly flows}: Number of outbound flows per hour
    \item \emph{Hourly unique destination IPs}: Number of unique destination IP addresses per hour, corresponding with the number of destinations communicated
    \item \emph{Hourly unique destination ports}: Number of unique destination ports per hour, corresponding with the number of protocols used
\end{itemize}

\begin{figure*}[t]
\begin{minipage}{.45\linewidth}
\centering
\subfloat[]{\label{subfig:corr_IN_PKTS_supports_adding_apps}\includegraphics[height=0.23\textheight, trim={0cm 0cm 0cm 0cm},clip]
{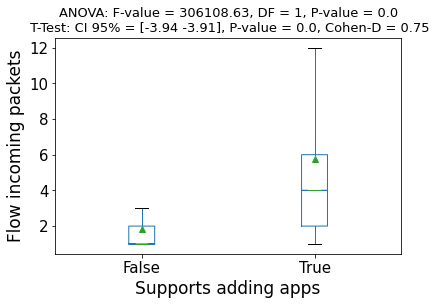}}
\end{minipage}
\hspace{\fill}
\begin{minipage}{.45\linewidth}
\centering
\subfloat[]{\label{subfig:corr_N_flows_per_hour_number_of_sensors}\includegraphics[height=0.23\textheight, trim={0cm 0cm 0cm 0cm},clip]
{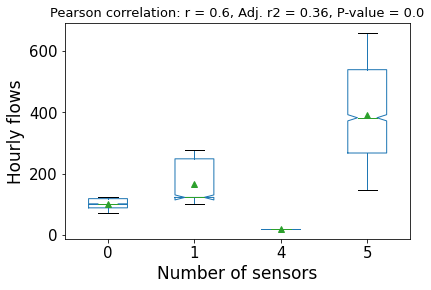}}
\end{minipage}
\vskip\baselineskip
\begin{minipage}{.45\linewidth}
\centering
\subfloat[]{\label{subfig:corr_N_distinct_IPV4_DST_ADDR_per_hour_number_of_actuators}\includegraphics[height=0.23\textheight, trim={0cm 0cm 0cm 0cm},clip]
{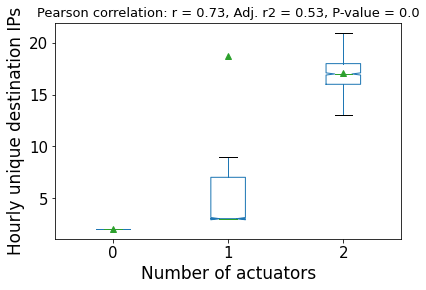}}
\end{minipage}
\hspace{\fill}
\begin{minipage}{.45\linewidth}
\centering
\subfloat[]{\label{subfig:corr_N_distinct_L4_DST_PORT_per_hour_CPU_speed_MHz}\includegraphics[height=0.23\textheight, trim={0cm 0cm 0cm 0cm},clip]
{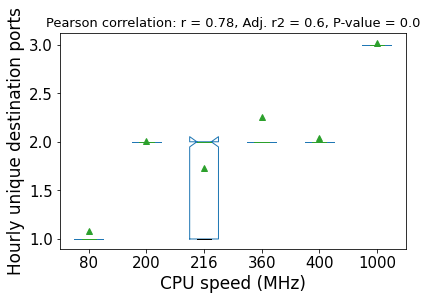}}
\end{minipage}
\caption{Correlation between IoT model complexity and network traffic predictability.}
\label{fig:complexity_predictability_correlation}
\end{figure*}

\subsection{Traffic Predictability of IoT vs. Non-IoT Devices}\label{subsec:Traffic_Predictability_of_IoT_vs_Non_IoT_Devices}

The basic reasoning which guides the notion of IoT traffic predictability is that IoT devices tend to have specialized functions with limited input, and they also normally have only a limited variety of functional states~\cite{Skowron2020}, such as ``On'' and ``Off'', ``Connecting to Wi-Fi'', ``idle'', ``status check'', etc., where each state has distinctive network traffic patterns. Thus, as opposed to typical non-IoT devices (e.g., servers, desktop computers, laptops, tablets and mobile phones~\cite{sivanathan2017characterizing, shahid2018}), IoT devices are hypothesized to behave in a relatively predictable manner in terms of traffic patterns. For instance, laptops and smartphones typically access a large number of web endpoints for browsing, while IoT devices normally send only automated pings (or messages with a predefined structure) to a finite number of endpoints.

In order to test the (reasonable) hypothesis regrading the relative predictability of IoT network traffic, compared with non-IoT devices, we \emph{(a)} conducted one-way ANOVA tests~\cite{bailey2008design} to ascertain a statistical difference of each KPI between the two groups (IoT vs. non-IoT), \emph{(b)} conducted two-sided t-tests~\cite{semenick1990tests} which provide 95\% confidence intervals (CIs) of the difference in the means of each KPI, and \emph{(c)} visualized the distributions of these KPIs using boxplots. As can be seen in Fig.~\ref{fig:iot_vs_non_iot}, the ANOVA and t-tests reveal significant differences ($P-value\leq0.05$) between IoT and non-IoT devices. That is, IoT devices tend to have fewer packets in an inbound flow as well as fewer total outbound flows per hour; IoT devices also communicate with fewer destinations than non-IoT devices while using fewer ports. As visualized by the boxplots in Fig.~\ref{fig:iot_vs_non_iot}, not only the means of the KPIs differ between IoT and non-IoT, but rather also the range and the variability of each KPI are smaller for the evaluated devices. For example, in most cases (see plot and CI in Fig~\ref{subfig:N_distinct_IPV4_DST_ADDR_per_hour_IoT_vs_Non_IoT}), IoT devices communicate with [20.35, 39.41] fewer destination IP addresses per hour, and in any case this number reaches only a few dozens at most (depicted as outliers). In comparison, non-IoT devices have demonstrated several hours, during which hundreds of destination IPs were communicated per hour.

\subsection{Variability in Traffic Predictability among IoT Models}\label{subsec:Variability_in_Traffic_Predictability_among_IoT_Models}

In many cases, the relative predictability of IoT network traffic makes it feasible to establish a solid baseline profile of normal activities. 
As such, it has been leveraged for several security-related applications such as device type identification~\cite{miettinen2017iot, meidan2017profiliot}, user authentication~\cite{shi2017smart, ashibani2018user}, human activity recognition~\cite{apthorpe2017smart, nweke2018deep} and anomaly detection for intrusion detection~\cite{meidan2018nbaiot, doshi2018machine}. Nevertheless, in the context of anomaly detection,
the authors of~\cite{meidan2018nbaiot, doshi2018machine, cvitic2020definition} suspected that not all of the IoT models have equally predictable network traffic patterns. As a result, certain IoT models are more amenable than others to network anomaly detection, even if the anomaly is due to the same reason (i.e., the same cyber-attack). 

To quantitatively investigate the above suspicion regarding unequality of traffic predictability among disparate IoT models, we conducted one-way ANOVA tests. As evident in Fig~\ref{fig:traffic_predictability_within_iot}, significant differences in means ($P-value=0.0$) were found, suggesting that dissimilar sub-populations (i.e., groups of IoT models) are present for each of the four KPIs. For instance, as illustrated in Fig.~\ref{subfig:N_distinct_IPV4_DST_ADDR_per_hour_within_IoT_only}, webcam/Amcrest/IPM-721W~\cite{amcrestipm721} communicates with fewer destination IP addresses per hour (and also with lower variability) than doorbell/Amazon/Ring~\cite{amazonring}. In terms of our use-case, this means that eventually, if these two IoT models are recruited to the same botnet and take part in executing a DDoS attack pointed at the same victim (identified by a destination IP address), this attack would likely pop up more clearly as an anomaly for webcam/Amcrest/IPM-721W (i.e., higher D-Score).

For quantitative investigation of traffic predictability, in addition to comparing KPI means using ANOVA tests (as described above), we also compare the KPIs' tendency to deviate from a random walk using the Hurst exponent~\cite{hurst1951long}, denoted as $H$. This technique has been widely used in a variety of domains, including finance~\cite{qian2004hurst}, medicine~\cite{subha2010eeg} and IoT security~\cite{dymora2019anomaly}. In practice, we treat the consecutive values of each KPI as a time series and then we calculate $H$ and use it to classify the KPI to one of three categories: \emph{Brownian motion} or \emph{geometric random walk} ($H=0.5$); \emph{anti-persistent behavior} ($H<0.5$), where the time series reverts to the mean; or \emph{persistent trending behavior} ($H>0.5$). Generally, the farther $H$ is from 0.5, the stronger non-random (i.e., the more predictable) the KPI's behavior is. To calculate $H$ we used a designated Python package named \emph{hurst}~\cite{hurstpackage} and applied it to the IoT-deNAT dataset~\cite{meidan_yair_2020_3924770}, which we previously used for analyzing traffic predictability. The $H$ values calculated for this dataset are presented in Table~\ref{tab:hurst}, where the rows correspond to the same devices in Table~\ref{tab:model_stats}, and the four right hand side columns correspond to the same KPIs in Figs.~\ref{fig:iot_vs_non_iot}, ~\ref{fig:traffic_predictability_within_iot} and~\ref{fig:complexity_predictability_correlation}. In each column, the $H$ values that are the closest or farthest from 0.5 (the point which indicates the least predictability) are highlighted in bold. With regard to the comparison of traffic predictability between IoT and non-IoT devices, as can be seen in Table~\ref{tab:hurst}, in most cases, IoT devices are the ones to have $H$ values farthest from 0.5, meaning that they are more predictable than non-IoT devices. This difference in $H$ is most prevalent in the `Flow incoming packets' KPI, where for IoT devices $H$ is very close to one (indicating that for this KPI a high value is often followed by an even higher one, and vice versa), while for the non-IoT devices $H$ is closer to 0.5 (indicating weaker connections between consecutive values). Regarding the comparison of traffic predictability among IoT models, Table~\ref{tab:hurst} presents non-negligible ranges of $H$. This finding led us to further investigate this variability and examine whether an IoT device's network traffic predictability is associated with its complexity, as described next.

\begin{table*}[ht]
 \centering
 \caption{Hurst exponent ($H$) values calculated for devices in the IoT-deNAT dataset~\cite{meidan_yair_2020_3924770}}
\label{tab:hurst}
\resizebox{\textwidth}{!}{%
\begin{tabular}{l|l|c|c|c|c}
\hhline{======}
\multicolumn{1}{c|}{\textbf{Category}} & \multicolumn{1}{c|}{\textbf{Model}} & \textbf{Flow incoming packets} & \textbf{Hourly unique destination IPs} & \textbf{Hourly unique destination ports} & \textbf{Hourly flows} \\ \hhline{======}
                                       & webcam/Amcrest/IPM-721W             & 1.001                          & \textbf{1.256}                         & 1.264                                    & 0.144                 \\ \cline{2-6} 
                                       & webcam/Samsung/SNH-1011N            & 0.988                          & \textbf{0.399}                         & 0.984                                    & 0.640                 \\ \cline{2-6} 
                                       & light\_bulb/TP\_Link/LB130          & \textbf{1.139}                 & 1.219                                  & \textbf{1.359}                           & 0.963                 \\ \cline{2-6} 
IoT                                    & socket/Wemo/Insight                 & 0.965                          & 0.842                                  & 0.949                                    & \textbf{0.277}        \\ \cline{2-6} 
                                       & speaker/Sonos/One                   & 0.932                          & 0.962                                  & 1.063                                    & 0.785                 \\ \cline{2-6} 
                                       & doorbell/Amazon/Ring                & 0.623                          & 0.975                                  & 1.056                                    & 0.724                 \\ \hhline{~=====} 
                                       & \textbf{Average}                             & 0.941                          & 0.942                                  & 1.113                                    & 0.589                 \\ \cline{2-6}
                                       & \textbf{Range}                  & 0.516                          & 0.857                                  & 0.410                                    & 0.819                 \\ \hhline{======}
                                       & laptop/Dell/Latitude\_E6430         & 0.665                          & 0.980                                  & 1.064                                    & \textbf{1.004}        \\ \cline{2-6} 
                                       & smartphone/Samsung/Galaxy\_Note\_5  & \textbf{0.602}                 & 0.951                                  & \textbf{0.855}                           & 0.944                 \\ \cline{2-6} 
Non-IoT                                & laptop/Dell/Latitude\_7400          & 0.837                          & 0.972                                  & 0.913                                    & 0.857                 \\ \hhline{~=====} 
                                       & \textbf{Average}                             & 0.701                          & 0.968                                  & 0.944                                    & 0.935                 \\ \cline{2-6}
                                       & \textbf{Range}                  & 0.235                          & 0.029                                  & 0.209                                    & 0.147                 \\ \hhline{======}
\end{tabular}%
}
\end{table*}

\subsection{Association of IoT model complexity with network traffic predictability}\label{subsec:association_of_IoT_model_complexity_with_network_traffic_predictability}

In this research, we hypothesize that the variability of traffic predictability among disparate IoT models (demonstrated above) is correlated with their difference in basic complexity, and can also be translated into varying levels of attack detectability. That is, the more complicated an IoT model is (in terms of hardware, software, user interaction, and networking), the less predictable its network traffic would be, such that generating a solid baseline profile of normal traffic behavior is less feasible. As a result, any attack would be less predictable by an AIDS, because anomalies would be harder to detect within noisy and non-deterministic traffic data. To test the hypothesis regarding the association between the complexity of an IoT model and its network traffic predictability, we characterized the evaluated IoT models based on features from our proposed taxonomy (introduced in Section~\ref{sec:taxonomy}) and correlated them with the four KPIs. As illustrated in Fig.~\ref{fig:complexity_predictability_correlation}, although not linear or entirely monotonous, trends do seem to exist between the static features (i.e., complexity characteristics) and the dynamic features (i.e., network traffic KPIs) of the evaluated IoT devices. For example, higher values of hourly flows are associated with a larger number of sensors (Fig.~\ref{subfig:corr_N_flows_per_hour_number_of_sensors}), logically because more data is captured from the IoT device's surroundings by multiple and various sensors, and the data is delivered to the designated servers via the Internet using larger quantities of outbound network traffic.

In this section we investigated the predictability of IoT network traffic from three complementary perspectives (IoT vs. non-IoT devices, differences among IoT models, and correlation with static IoT features). As part of this investigation, we conducted statistical hypothesis testing and visualization using data from~\cite{meidan2018nbaiot}. As can be seen in Appendix~\ref{apndx:UNSW}, we obtained similar findings from another freely-available dataset~\footnote{https://iotanalytics.unsw.edu.au/iottraces}, collected by UNSW researchers~\cite{Sivanathan2019classifying} from various IoT and non-IoT devices during 20 days. Some of our experimental findings, obtained using the IoT-deNAT and the UNSW datasets, conform with past research, as follows: Regarding the traffic predictability of IoT vs. non-IoT devices (Subsection~\ref{subsec:Traffic_Predictability_of_IoT_vs_Non_IoT_Devices}), similar to us the authors of~\cite{sivanathan2017characterizing} noted that overall, IoT devices generate much less traffic compared to traditional non-IoT devices. Also, according to their findings, an IoT device typically transfers only less than 1 KB per session. However, since their goal was to facilitate IoT model fingerprinting, their quantitative evaluation focused on finding distinct levels of traffic attributes which are characteristic of certain IoT models, rather than on  statistical validation of the \emph{differences} between IoT and non-IoT (as in our case). In a later paper~\cite{Sivanathan2019classifying} which extended the above-mentioned fingerprinting goal and methodology, the results also highlighted the most prominent values of traffic features (e.g., distinctive destination ports) instead of comparing the number of unique values per hour between IoT and non-IoT (like we do). Regarding the variability in traffic predictability among IoT models (Subsection~\ref{subsec:Variability_in_Traffic_Predictability_among_IoT_Models}), the authors of~\cite{Majumdar2020IATstats} found statistically significant differences in the distribution of packet inter-arrival time among disparate IoT models, and the authors of~\cite{Skowron2020} visualized the distribution of packet size among several IoT devices. However, each of these studies focused on one feature only, whereas we statistically compared among IoT models using four other (dynamic) features, while also correlating them with (static) device features. The authors of~\cite{cvitic2020definition} classified the traffic predictability of smart home IoT devices into four levels, using the coefficient of variation of the ratio of the received and sent volume of traffic. However, as opposed to our study, they did not compare the IoT predictability with non-IoT, nor did they associate traffic predictability with device complexity or attack detectability. Regarding the association of IoT model complexity with network traffic predictability (Subsection~\ref{subsec:association_of_IoT_model_complexity_with_network_traffic_predictability}), we were unable to find any existing study which performed a similar analysis.

\begin{figure}[h]
\centerline{\includegraphics[width=1.0\columnwidth, trim={0 4.5cm 7cm 0},clip]{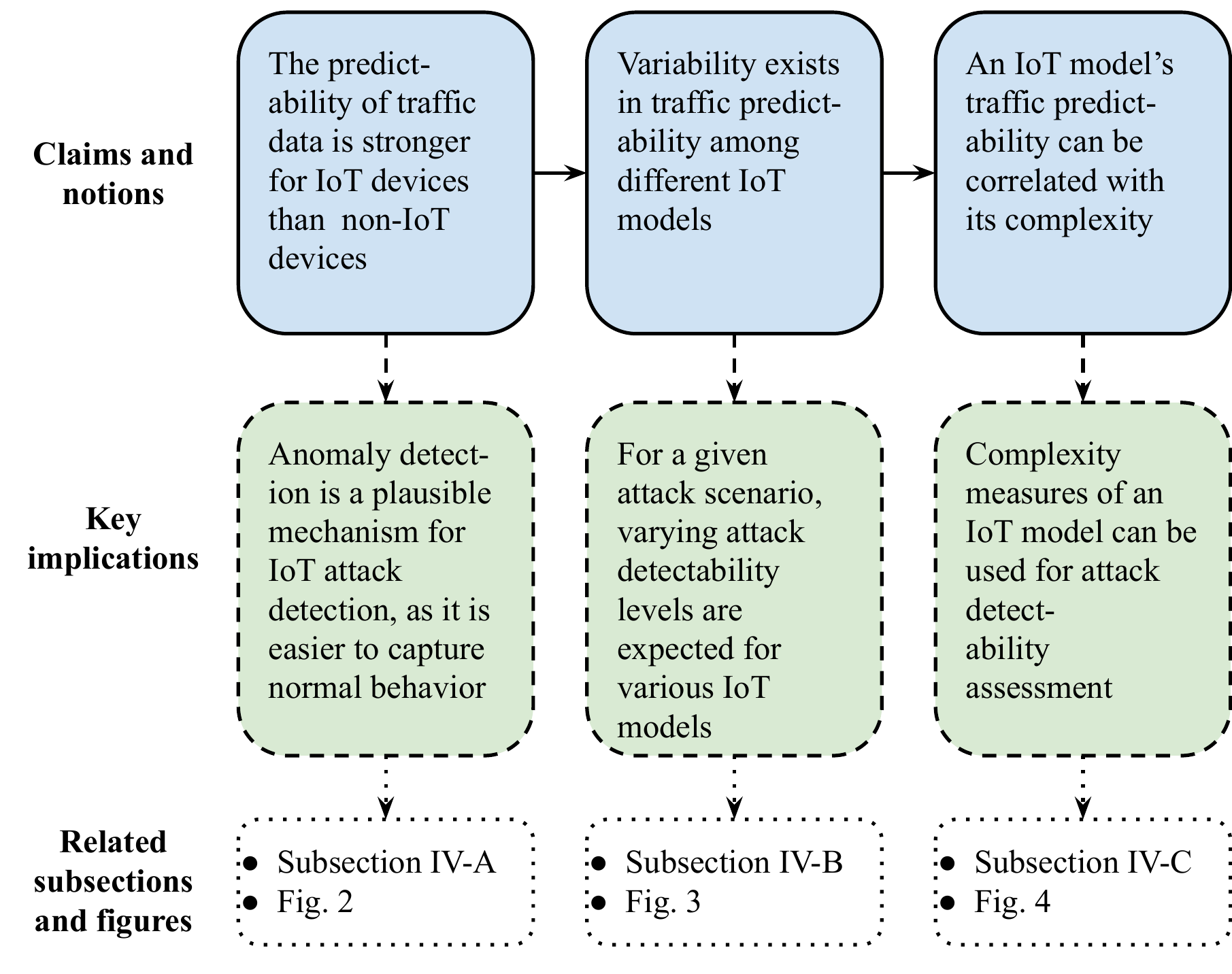}}
\caption{The logic flow behind the statistical analysis covered in Section~\ref{sec:iot_traffic_predictability}, which motivated the development of our IoT attack detectability assessment method (D-Score).}
\label{fig:stat_logic}
\end{figure}

\begin{figure*}[ht]
\centerline{\includegraphics[width=0.95\textwidth, trim={0 9.2cm 0 6.6cm},clip]{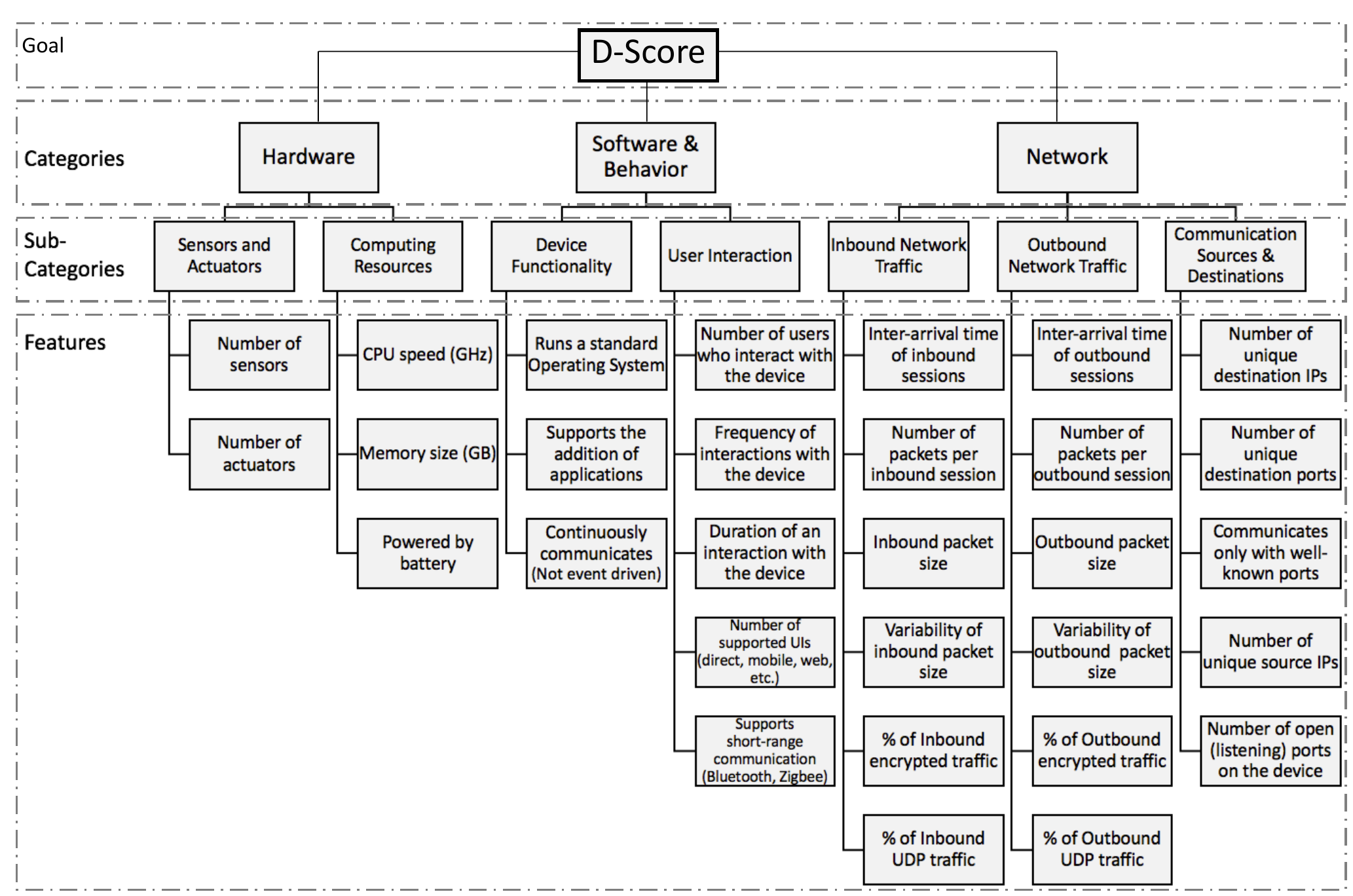}}
\caption{Proposed taxonomy for IoT attack detectability.}
\label{fig:taxonomy}
\end{figure*}

The logic flow (illustrated in Fig.~\ref{fig:stat_logic}) behind the statistical analysis covered in this section motivated the development of our IoT attack detectability assessment method (D-Score). That is, if IoT network traffic is more predictable than non-IoT traffic, variability exists in traffic predictability among IoT models, and this variability can be correlated with the IoT model's complexity, then anomaly-based attack detection is a plausible mechanism for attack detection in the IoT, where features that capture the complexity of an IoT model can be utilized to assess its attack detectability, i.e., its D-Score. Given this motivation, in the sections that follow we \emph{(a)} present a novel taxonomy which is designed to characterize the basic complexity of a given IoT model (namely, ``static features'') as well as the predictability of its network traffic (namely, ``dynamic features''), and \emph{(b)} propose and empirically evaluate an expert based method which assigns weights to those features in the context of a given attack scenario, in order to ultimately produce a risk assessment in the form of D-Score.

\section{Taxonomy for IoT Attack Detectability}\label{sec:taxonomy}

The authors of~\cite{kovanen2016survey} propose a concise list they refer to as \emph{detectable attack features}. It consists of twelve network traffic features to be analyzed by AIDSs, since they are the ones most likely to change upon the occurrence of a cyber-attack. In contrast, we propose a considerably richer taxonomy (illustrated in Fig.~\ref{fig:taxonomy}), which consists of thirty \emph{features}, hierarchically distributed among three \emph{categories} and seven \emph{sub-categories}. Of the three categories in our taxonomy, the \emph{Network} is largely congruent with the above list of detectable attack features. 
We refer to them as \emph{direct features}, since they are extracted from network traffic and directly analyzed by an AIDS. The direct features in our taxonomy have been carefully designed by domain experts using a systematic process, following a thorough review of past related research, e.g.,~\cite{meidan2018nbaiot, doshi2018machine, kovanen2016survey, shaikh2019iot,
brun2018iot, Lin2014Botnet}.

\begin{table*}[ht]
\caption{Categories, sub-categories and features in our proposed taxonomy, as well as the means to evaluate them.}\label{tab:taxonomy_features}
\centering
\resizebox{\textwidth}{!}{%
\begin{tabular}{l|l|l|l|l}
\hhline{=====}
\multicolumn{1}{c|}{\emph{\textbf{Category}}} &
  \multicolumn{1}{c|}{\emph{\textbf{Sub-category}}} &
  \multicolumn{1}{c|}{\emph{\textbf{Feature}}} &
  \multicolumn{1}{c|}{\emph{\textbf{Data Source / Calculation / Description}}} &
  \multicolumn{1}{c}{\emph{\textbf{Unit}}} \\ 
\hhline{=====}
\multirow{5}{*}{\makecell{HW: \\ Hardware}} &
  \multirow{2}{*}{\makecell{SNA: \\ Sensors \& actuators}} &
  NSNS: Number of sensors &
  Spec sheet &
  Count \\ \cline{3-5} 
 &
   &
  NACT: Number of actuators &
  Spec sheet &
  Count \\ \cline{2-5} 
 &
  \multirow{3}{*}{\makecell{RSR: \\ Computing resources}} &
  CPUS: CPU speed &
  Spec sheet &
  MHz \\ \cline{3-5} 
 &
   &
  MEMS: Memory size &
  Spec sheet &
  MB \\ \cline{3-5} 
 &
   &
  BATT: Powered by battery &
  Spec sheet &
  1/0 \\ 
\hhline{=====}
\multirow{8}{*}{\makecell{SB: \\ Software \& \\ behavior}} &
  \multirow{3}{*}{\makecell{FNC: \\ Device functionality}} &
  STOS: Runs a standard operating system &
  nmap -O \textless{}IP address\textgreater{} &
  1/0 \\ \cline{3-5} 
 &
   &
  ADAP: Supports the addition of applications &
  Spec sheet &
  1/0 \\ \cline{3-5} 
 &
   &
  CCOM: Continuously communicates (not event driven) &
  (Mean hourly packets at night) / (max hourly packets during the day) &
  \% \\ \cline{2-5} 
 &
  \multirow{5}{*}{\makecell{INT: \\ User interaction}} &
  NUSR: Number of users who interact with the device &
  Expected number of unique users &
  Count \\ \cline{3-5} 
 &
   &
  FINT: Frequency of interactions with the device &
  Expected mean number of hourly user interactions &
  Count/hour \\ \cline{3-5} 
 &
   &
  DINT: Duration of an interactions with the device &
  Expected mean duration &
  Seconds \\ \cline{3-5} 
 &
   &
  NUIS: Number of supported UIs (direct, mobile, web, etc.) &
  Number of unique user interfaces &
  Count \\ \cline{3-5} 
 &
   &
  SRNG: Supports short-range communication (Bluetooth, Zigbee) &
  Spec sheet &
  1/0 \\ 
\hhline{=====}
\multirow{17}{*}{\makecell{NT: \\ Network}} &
  \multirow{6}{*}{\makecell{INB: \\ Inbound \\ network traffic}} &
  IATI: Inter-arrival time of inbound sessions &
  Median &
  Seconds \\ \cline{3-5} 
 &
   &
  PCKI: Number of packets per inbound session &
  Median &
  Count \\ \cline{3-5} 
 &
   &
  PCSI: Inbound packet size &
  Sum(inbound packet sizes) / count(inbound sessions) &
  Bytes \\ \cline{3-5} 
 &
   &
  PCVI: Variability of inbound packet size &
  St.Dev(inbound packet size) &
  Bytes \\ \cline{3-5} 
 &
   &
  ENCI: \% of inbound encrypted traffic &
  Count(inbound packets to ports 443 or 8443) / count(inbound packets) &
  \% \\ \cline{3-5} 
 &
   &
  UDPI: \% of inbound UDP traffic &
  Count(inbound UDP packets) / count(inbound packets) &
  \% \\ \cline{2-5} 
 &
  \multirow{6}{*}{\makecell{OUT: \\ Outbound \\ network traffic}} &
  IATO: Inter-arrival time of outbound sessions &
  Median &
  Seconds \\ \cline{3-5} 
 &
   &
  PCKO: Number of packets per outbound session &
  Median &
  Count \\ \cline{3-5} 
 &
   &
  PCSO: Outbound packet size &
  Sum(outbound packet sizes) / count(outbound sessions) &
  Bytes \\ \cline{3-5} 
 &
   &
  PCVO: Variability of outbound packet size &
  St.Dev(outbound packet size) &
  Bytes \\ \cline{3-5} 
 &
   &
  ENCO: \% of outbound encrypted traffic &
  Count(outbound packets to ports 443 or 8443) / count(outbound packets) &
  \% \\ \cline{3-5} 
 &
   &
  UDPO: \% of outbound UDP traffic &
  Count(outbound UDP packets) / count(outbound packets) &
  \% \\ \cline{2-5} 
 &
  \multirow{5}{*}{\makecell{SRD: \\ Communication \\ sources \& destinations}} &
  DSIP: Number of unique destination IP addresses &
  Hourly median &
  Count \\ \cline{3-5} 
 &
   &
  DSPR: Number of unique destination port numbers &
  Hourly median &
  Count \\ \cline{3-5} 
 &
   &
  WLPR: Communicates only with well-known ports &
  Count(outbound packets to ports\textless{}=1024) / count(outbound packets) &
  1/0 \\ \cline{3-5} 
 &
   &
  SRIP: Number of unique source IP addresses &
  Hourly median &
  Count \\ \cline{3-5} 
 &
   &
  OPPR: Number of open (listening) ports on the device &
  nmap \textless{}IP address\textgreater{} &
  Count \\ 
\hhline{=====}
\end{tabular}%
}
\end{table*}

In addition to the direct features, for the quantification of IoT attack detectability, we also propose the novel use of \emph{indirect features}, which are those features that are not directly analyzed by an AIDS, but are likely to affect the direct features. The indirect features mainly fall in the \emph{Hardware} and \emph{Software and Behavior} categories, (see Table~\ref{tab:taxonomy_features}). The indirect features are based mostly on domain knowledge as well as reasoning and experimentation concerning the correlation between device complexity and traffic predictability (see Subsection~\ref{subsec:association_of_IoT_model_complexity_with_network_traffic_predictability}). For instance, an indirect feature that we find to be potentially indicative of IoT cyber-attacks is the \emph{Number of sensors}. This feature is not analyzed directly by an AIDS, however, it is likely to affect direct features such as the \emph{Variability of outbound packet size} or the number of \emph{Hourly flows} (see Fig.~\ref{subfig:corr_N_flows_per_hour_number_of_sensors}). For example, an IoT device with just one sensor (like a smart motion detector) is likely to have much more stable and predictable patterns of outbound traffic than another device which also includes a webcam, microphone, temperature sensor, humidity sensor, and ambient light sensor (like a smart baby monitor). This is based on the logic that more sensors mean more communication protocols and therefore, greater diversity of packet sizes. The number of sensors can also affect the number of destination ports, destination IP addresses, variability of session lengths and inter-arrival times. Altogether, when the complexity of an IoT device is relatively low (e.g., it uses only a few sensors), the resultant traffic patterns are expected to be rather predictable and facilitate the creation of a normal behavior profile. In turn, traffic anomalies due to cyber-attacks are more likely to be prominent, and the IoT attack detectability (i.e., D-Score) is expected to be higher.

When designing the taxonomy we balanced a number of considerations, such that the taxonomy would be:

\begin{itemize}[leftmargin=*]         
\item \textbf{IoT orientated}: As opposed to the general purpose OSI models discussed in Subsection~\ref{subsec:existing_taxonomies}, our taxonomy is more specific to the IoT. For example, the features under the \emph{Sensors and actuators} sub-category might not be relevant for PCs. Additionally, features such as the \emph{Number of unique destination IPs} are more informative for IoT devices than for PCs, because unlike PCs, IoT devices typically communicate with only a small set of predefined servers.
\item \textbf{Extensive}: Capable of (hierarchically) describing any IoT device from a broad range of perspectives, without leaving out aspects which are potentially informative for attack detection. These include hardware, software, user interaction, and networking viewpoints for any IoT model.
\item \textbf{Informative for attack detection}: This includes direct features which are likely to behave abnormally in case of an attack as well as indirect features which affect the values and stability of the direct features.
\item \textbf{Flexible}: Able to facilitate D-Score computation for \emph{(a)} a large variety of known and unknown attack scenarios, and \emph{(b)} unlimited IoT models (within the scope defined in Subsection~\ref{subsec:scope}), since it does not depend on predefined hardware components or specific communication protocols.
\item \textbf{Coherent / self-explanatory}: Easily understood by security professionals with differing expertise areas, so they could thoughtfully respond to the questionnaire that uses this taxonomy. 
\item \textbf{Feasible}: Capable of easily capturing the features' values in practice, such that the D-Score computation for any given IoT model is achievable within a reasonable investment of time and effort.
\end{itemize}

Following is a description of our taxonomy's key components:

\subsection{HW: Hardware}\label{subsec:taxonomy_hardware}

\subsubsection{SNA: Sensors and Actuators}
\label{subsubsec:sensors_and_actuators}

Sensors and actuators are key characteristics of an IoT device, which typically interacts with the physical world by collecting information from the device's surrounding environment and performing required actions~\cite{nawir2016iot}, as opposed to typical servers, PCs, and mobile devices. Some common sensors in IoT devices are motion detectors and temperature sensors, while lights and speakers are examples of IoT actuators. The \emph{total} number of sensors and actuators mainly affects the frequency of communication, and thus affects the inter-arrival time (IAT) of packets as well. 
Alternatively, the number of \emph{unique} sensors or actuators mainly affects the variability of packet sizes and the diversity of communication protocols, sources, and destinations (IP addresses and ports). 
However, for consumer IoT devices (unlike smart cars, for example), it is anticipated that the number of unique sensors and actuators should be almost identical to the total number of sensors and actuators, so for simplicity, we did not allocate distinct (and probably redundant) features for both.

\subsubsection{RSR: Computing Resources}\label{subsubsec:computing_resources}

The CPU, memory and power source are fundamental resources that can have an impact on packet traffic throughput~\cite{capacity2005}. For instance, an insufficient CPU would not allow a highly intensive flooding attack, and might lead to crashes and significant traffic abnormalities under such circumstances. Also, additional memory may better support the computation needed to encrypt network traffic. Battery-powered IoT devices (e.g., smart smoke detectors), for the most part, are associated with low computational requirements. This implies simple functionality with minimal network traffic volume, so that attack-related traffic abnormalities are likely to emerge.

\subsection{SB: Software and Behavior}\label{subsec:taxonomy_Software_and_behavior}

\subsubsection{FNC: Device Functionality}\label{subsubsec:device_functionality}

Non-standard IoT operation systems (OSs), defined here as any OS other than the general-purpose Linux OS, traditionally used in high-end IoT devices~\cite{Hahm2016linux}, are considerably more task-oriented. Therefore, in view of attack detectability, the normal traffic patterns of IoT devices which run these task-oriented lightweight OSs should be more predictable and thus, abnormalities should stand out more clearly. In addition, if an IoT device supports the addition of apps (as in smart TVs and smartwatches), a larger variety of communication servers, protocols, packet sizes, and IATs is foreseen, thus traffic predictability is likely to decrease. In contrast to IoT devices which produce network traffic very rarely, sometimes triggered only upon predefined scenarios such as the detection of motion or smoke, other devices communicate more frequently, e.g., for activity monitoring, tracking temperature or humidity, and so forth. Rationally, the less continuous normal traffic is, the more prominent any attack-related anomaly is expected to be.

\subsubsection{INT: User Interaction}\label{subsubsec:user_interaction}

The user interaction (UA) features are ``softer'' and less technical. Nevertheless, as UA is strongly associated with traffic generation~\cite{apthorpe2019keeping, subahi2019detecting}, we find that it is a fundamental component of the taxonomy. In this regard, more users logically means more usage and traffic patterns and thus more variability in the detectable attack features~\cite{kovanen2016survey}. For instance, the typically short~\cite{doshi2018machine} packet IAT in flooding attacks as part of a DDoS is less likely to come through as a traffic anomaly when more variability is present in the packet IAT, due to multiple users and UA patterns. Additionally, for IoT devices which can be interacted with not only via an app but rather also directly (e.g., a smartwatch) and/or via a web application (e.g., a smart webcam), more traffic variability is expected to be found and obscure attack-related traffic anomalies.

\subsection{NT: Network}\label{subsec:taxonomy_network}

These \emph{direct features} resemble the previously mentioned \emph{detectable attack features}~\cite{kovanen2016survey}, and they are calculated using captured traffic data.

\subsubsection{INB and OUT: Inbound and Outbound Network Traffic}\label{subsubsec:inbound_outbound_network_traffic}

Among the direct/detectable features, inbound network traffic features describe the traffic whose destination is the IoT device in terms of packet count and size, encryption, etc. The outbound network traffic features are similar to the inbound ones, however in this case, the IoT device is the source of the traffic for the features rather than the destination. In past research, e.g.,~\cite{meidan2018nbaiot, kovanen2016survey, shaikh2019iot, khraisat2019novel}, these features were found informative for IoT attack detection. For instance, most normal IoT packets are sent at regular time intervals~\cite{Majumdar2020IATstats} (e.g., automated network activities). In contrast, these intervals are dramatically shorter~\cite{doshi2018machine} for most DoS attack traffic and thus informative. The authors of~\cite{doshi2018machine} also observed that as opposed to the size variations of normal traffic, TCP SYN flood packets are quite small (typically under 100 bytes), largely in order to maximize the number of connection requests per second for victim exhaustion.

\subsubsection{SRD: Communication Sources \& Destinations}\label{subsubsec:communication_sources_and_destinations}

As opposed to PCs and smartphones, IoT devices normally communicate with a limited number of endpoints (IP destination)~\cite{Heimdall2017IP}, e.g., for (de)activation from the cloud, retrieving firmware updates, and logging their status; thus, an increased number of destination endpoints might be indicative of attack traffic (e.g., sending messages to the botnet victims), while extraneous source IP addresses might indicate logging attempts into the device, as in Mirai~\cite{antonakakis2017understanding}. Moreover, the set of destination IPs rarely changes over time~\cite{doshi2018machine}. Also, the traffic associated with DDoS attacks has different protocol distributions (roughly equivalent to the number of destination ports), and they also include fewer total protocols~\cite{doshi2018machine}, thus highly likely to be informative for IoT attack detection.

\section{Proposed Method for Computing the D-Score of an IoT Model for an Attack Scenario}\label{sec:proposed_method}

Our taxonomy includes various features that are potentially relevant, either directly or indirectly, to the ability of an AIDS to detect attacks which involve IoT devices. Understanding the importance of those features to the D-Score of various attack scenarios is a crucial task, as the contribution of features may not be the same for each attack scenario. Two main approaches can be considered for this kind of problem; one is data-driven and the other is expert-based. In order to reduce the considerable investment associated with the former approach (see Subsection~\ref{subsec:expert_vs_data}) and still obtain a robust and highly generalizable detectability assessment mechanism, we suggest using an expert-based approach. As part of the suggested method, cyber-security experts rank the taxonomy's features with respect to attack scenarios of interest, with the aim of producing feature weights. However, ranking multiple criteria simultaneously is a difficult task for an individual to accomplish, often resulting in an inaccurate ranking~\cite{saaty2008decision}. The \emph{analytical hierarchical process (AHP)} method was introduced by Saaty~\cite{saaty1988analytic} to moderate this drawback. AHP represents the problem space as a hierarchical structure in which the experts only compare pairs of elements at the same level of the hierarchy, thus reducing the amount of comparison needed to a more reasonable amount and resulting in a more precise ranking. The number of pairwise comparisons among $n$ elements in a given level using AHP is calculated as $\frac{n \cdot (n-1)}{2}$. Based on the pairwise comparisons done by each expert, the AHP computes a vector of feature weights~\cite{shih2003method}, which is eventually averaged over all the experts.

\begin{figure}[h!]
\centerline{\includegraphics[width=1.0\linewidth, trim={0 1.5cm 4.7cm 0},clip]{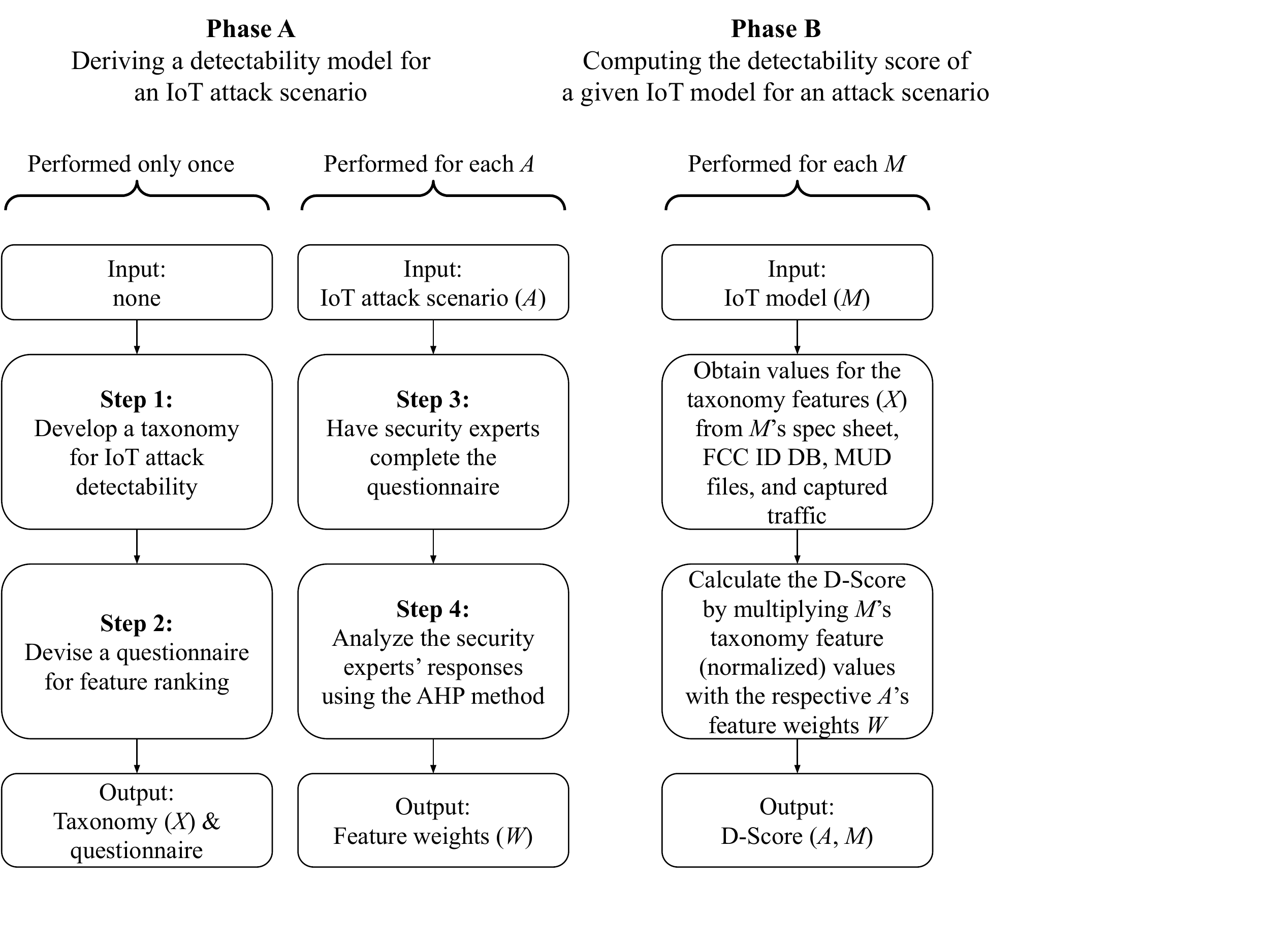}}
\caption{Flow chart of the key steps in the two phases of our proposed method.}
\label{fig:method_step_by_step}
\end{figure}

Essentially, our method for quantifying IoT attack detectability includes two key phases which conform with AHP's requirements:
\begin{description}[style=unboxed,leftmargin=0cm]
\item[\textbf{Phase A:}]
Deriving a detectability model for an IoT attack scenario $A$. This model reflects the contribution (weight, denoted as $W$) of each feature from the taxonomy to the detectability of the corresponding attack scenario. Note that the weights $W$ are calculated \emph{only once for a given attack scenario $A$}, and can later be applied to multiple IoT models $M$s in order to assess the detectablity of $A$ on each of them. 
\item[\textbf{Phase B:}]
Computing the detectability score of a given IoT model $M$ for an attack scenario $A$, using the detectability model derived in phase A. This is done by evaluating the relevant features (denoted as $X$) for the given IoT model, and using the feature weights to calculate the D-Score as a weighted sum ($W \cdot X$).  
\end{description}

Once phase B is complete, if the calculated D-Score for an attack is low (i.e., it would be hard to detect the attack) and yet the organization decides to deploy the IoT device, proper countermeasures are encouraged to be taken in order to mitigate cyber-risks. These can include~\cite{Bertino2017BotnetsSecurity} \emph{(a}) changing default passwords, \emph{(b)} installing security patches, \emph{(c}) disabling UPnP, \emph{(d}) monitoring certain TCP/IP ports (e.g., 23 or 2323 to defend from attempts to gain unauthorized control via Telnet), \emph{(e)} implementing additional network isolation, \emph{(f)} updating firewall rules, etc. Further countermeasures can be found in, e.g.,~\cite{8895034, Abdul2017countermeasures}, from which the organization should implement the ones most relevant to the anticipated attack. 

Our proposed method is envisioned as a means of facilitating risk assessment and mitigation processes, and support IoT deployment decisions which are often made by a wide range of users (from home users to enterprise network administrators). When using our method, a simple comparison between the D-Score labels (Fig.~\ref{subfig:label_example}) of competing IoT models is all that is needed, similar to the manner in which a consumer would compare energy efficiency rating labels before purchasing an appliance. However, in order to associate an appropriate D-Score label with an IoT model, we advise implementing the two phases of our proposed method, as outlined above and illustrated in Fig.~\ref{fig:method_step_by_step}. Phase A, the research-oriented phase, can be performed by a specialized cyber-security research institute. In comparison, phase B, which is more technical, should be performed for every newly released IoT model by a central standards institution, in order to ensure the uniformity and trustworthiness of the D-Score labels. Further details about the steps comprising the two phases of our method are provided in the subsections that follow; a step-by-step quantitative demonstration of our proposed method is provided in Section~\ref{sec:evaluation_method_and_validation}.

\subsection{Deriving a Detectability Model for an IoT Attack Scenario}\label{subsec:deriving_a_detectability_model_for_an_iot_attack_scenario}

The first phase of our method consists of the following four steps, where steps 1 and 2 are generic (i.e., performed and fine-tuned only at the beginning), while steps 3 and 4 are performed for each $A$ separately.

\begin{enumerate}[leftmargin=*]
    \item \textbf{Developing a taxonomy for IoT attack detectability}: The taxonomy (described in Section~\ref{sec:taxonomy}) is a hierarchical decomposition of the attributes of IoT models into a list of categories, sub-categories and features. We designed our taxonomy to be generic so it could support a variety of IoT models and attack scenarios.
    \item \textbf{Devising a respective questionnaire used for feature ranking}: The questionnaire is comprised of     questions for cyber-security experts, where each question compares two elements (from the taxonomy) in terms of their relevance and relative importance to detecting a specific IoT attack scenario. For D-Score's taxonomy, the default AHP implementation would require the experts to compare three pairs of categories (i.e., $\frac{3 \cdot (3-1)}{2}$), five pairs of sub-categories ($1+1+3$), and also 57 pairs of features ($1+3+3+10+15+15+10$), resulting in a total of 65 comparisons. Although less than the number of all pairwise comparisons required in a naïve approach, it still requires a significant amount of expert time. Thus, to further reduce the number of comparisons, we propose the following process:
\begin{itemize}[leftmargin=*]
    \item \emph{Preliminary filtering of categories:} The categories are not assigned weights, but are only used by the experts to focus their attention on the sub-categories and features that really matter for attack detection. As discussed in Section~\ref{subsec:questionnaire}, this novel procedure was used broadly by the experts in our study and saved effort on their part and much computation. 
    \item \emph{Preliminary filtering of sub-categories in the selected categories:} Irrelevant sub-categories may remain within a potentially informative category (i.e., not filtered out), so we provide the ability to filter them out as well, prior to the comparison among sub-categories and features.
    \item \emph{Pairwise comparisons among sub-categories:} In order to assign weights to the remaining sub-categories, each expert is required to rank pairs of sub-categories. 
    \item \emph{Pairwise comparisons among the features in each sub-category}: Similarly, pairwise comparison of features is required, however only in the sub-categories previously deemed as relevant by the expert.
\end{itemize}

    \item \textbf{Having security experts answer the questionnaire}: For a given attack scenario $A$ (e.g., DDoS flooding), cyber-security researchers and practitioners who are familiar with $A$ are requested to use their experience and best judgment to respond to the pairwise comparisons in the questionnaire.
    \item \textbf{Analysing the security experts' responses using the AHP method to assign feature weights}: The experts' responses are analyzed using AHP, resulting in a weight assigned to each feature of the taxonomy. This weight is the average of the respective weights provided by all of the experts who completed the questionnaire for the given attack scenario $A$. The (averaged) feature weights $W$ determine: \emph{How informative each feature is for detecting the selected attack scenario on any given IoT model? Which are the most informative features for each attack scenario?}
\end{enumerate}

A challenge arises when the features $X$ do not share the same range of values (see Table~\ref{tab:alpha_derivation}). To overcome this challenge, feature normalization was used in prior research by, e.g., expert-based ranking~\cite{saaty2008decision} (i.e., another level at the bottom in the taxonomy's hierarchy), manual division into equivalent groups~\cite{saaty1988analytic, saardchom2012credit}, or automatic binning~\cite{yean2014relative}. In contrast, for normalization we propose using the hyperbolic tangent (tanh) function with some fine-tuning, as it is \emph{(a)} systematic and consistent, \emph{(b)} requires less human intervention, \emph{(c)} is non-linear and \emph{(d)} its slope can be easily adjusted. We use the following notation to describe our proposed method for feature normalization:

\begin{description}[style=unboxed,leftmargin=0cm]
\item[\bm{$i\in[1, |features|]$}:] Index of a feature in the taxonomy
\item[\bm{$j\in[1, |attacks|]$}:] Index of an attack scenario of interest
\item[\bm{$x_i$}:] Original value of the $i^{th}$ feature
\item[\bm{$x_i^{min}, x_i^{max}$}:] Minimum and maximum values, respectively, in the expected range of $x_i$
\item[\bm{$x'_{ij}\in[0, 1]$}:] Normalized value of $x_i$ for attack scenario $j$
\item[\bm{$\alpha_{ij}\in\mathbb{R}$}:] Coefficient to control the shape of tanh function
\item[\bm{$\beta_{ij}\in\mathbb{R}$}:] Coefficient to control the value of $\alpha_{ij}$
\item[\bm{$\delta_{ij}\in\{-1, 1\}$}:] Direction of influence on a given attack scenario $j$ for increasing values of $x_i$
\item[\bm{$w_{ij}$}:] Weight assigned to feature $i$ and attack $j$ using AHP
\end{description}

To normalize the original value of $x_i\in[x_i^{min}, x_i^{max}]$ into $x'_{ij}\in[0, 1]$, first we calculate $\alpha_{ij}$ as in Eq.~\eqref{eq:alpha_calculation}, such that the range of $x_i$ is taken into account. For example, $x_i^{max}$ of percentage and binary features (e.g., ENCI and BATT, respectively) is $1=\log_{10}(0)$, the number of sensors (NSNS) is typically lower than $10=\log_{10}(1)$, the number of packets per inbound session (PCKI) is typically lower than $100=\log_{10}(2)$, etc. 

\begin{equation}
\alpha_{ij} = \frac{1}{\beta_{ij} \cdot 10^{\{\log_{10}(x_i^{max})-1\}}} \label{eq:alpha_calculation}
\end{equation}

Table~\ref{tab:alpha_derivation} outlines the feature ranges which are relevant for our research and the associated $\alpha$ values, calculated using Eq.~\eqref{eq:alpha_calculation}. 
For simplicity, we used $\beta=5$ for all of the feature ranges, so that each range of $x_i$s (i.e., a row in this table) has its $\alpha$.

\begin{table}[h]
\caption{Derived $\alpha$ for varied feature value ranges ($\beta=5$).}\label{tab:alpha_derivation}
\centering
\resizebox{\columnwidth}{!}{%
\begin{tabular}{c|c|c|c|c}
\hhline{=====}
\bm{$x_{min}$} & 
\multicolumn{1}{c|}{\bm{$x_{max}$}} & 
\bm{$Log_{10}(x_{max})$} & 
\multicolumn{1}{c|}{\bm{$\alpha$}}  &
\multicolumn{1}{|c}{\emph{\textbf{Features}}}\\ 
\hhline{=====}
0 & 1 & 0 & 2 & ENCI, BATT\\ \hline
0 & 10 & 1 & 0.2 & NSNS, FINT\\ \hline
0 & 100 & 2 & 0.02 & PCKI, IATO\\ \hline
0 & 1,000 & 3 & 0.002 & CPUS, IATI\\ \hhline{=====}
\end{tabular}
}
\end{table}

\begin{table*}[t]
 \centering
 \caption{Evaluated IoT attack scenarios}
\label{tab:attack_scenarios}
\resizebox{\textwidth}{!}{%
\begin{tabular}{
l|l|c|c
}
\hhline{====}
\multicolumn{1}{c|}{\emph{\textbf{Abbreviation}}} & \multicolumn{1}{c|}{\emph{\textbf{Description of the IoT attack scenario}}} & 
\multicolumn{1}{c|}{\emph{\textbf{Ref.}}} & \multicolumn{1}{c}{\emph{\textbf{Responses}}} \\
\hhline{====}
C\&C communication & \begin{tabular}[l]{@{}l@{}}The IoT device is communicating with a C\&C server\end{tabular}  & \cite{hallman2017ioddos} & 10 \\
\hline
DDoS flooding & The IoT device is executing a flood attack as part of a DDoS campaign & \cite{hallman2017ioddos} & 13 \\
\hline
Data exfiltration & The IoT device is exfiltrating data & \cite{d2016data} & 12 \\
\hline
Bot scanning & The IoT device is being scanned / brute-forced by a bot (as used by Mirai for propagation) & \cite{Bertino2017BotnetsSecurity} & 5 \\
\hhline{====}
\end{tabular}%
}
\end{table*}

After deriving the $\alpha$ values we determine $\delta_{ij}$, as increasing values of a feature may increase the D-Score for a given attack scenario (and then $\delta_{ij}=1$), but decrease it for others ($\delta_{ij}=-1$). Finally, for each original $x_i$ we calculate the normalized value $x'_{ij}$, as in Eq.~\eqref{eq:x_tag_calculation}.

\begin{equation}
x'_{ij} = \begin{cases} & tanh(\alpha_{ij} \cdot \delta_{ij} \cdot x_i) \text{ if } \delta_{ij} = 1 \\ & 1+tanh(\alpha_{ij} \cdot \delta_{ij} \cdot x_i) \text{ if } \delta_{ij} = -1 \end{cases}
 \label{eq:x_tag_calculation}
\end{equation}

\begin{table}[!b]
\caption{Percentage of respondents who kept each category and sub-category during the preliminary filtering process.}
\label{tab:preliminary_filtering_percentage}
\centering
\resizebox{\linewidth}{!}{%
\begin{tabular}{l|c|c|c|c|c|c|c|c|c|c}
\hhline{===========}
\emph{\textbf{}} & \multicolumn{3}{c|}{\emph{\textbf{Categories}}} & \multicolumn{7}{c}{\emph{\textbf{Sub-categories}}} \\ \hline
\multicolumn{1}{c|}{\emph{\textbf{Attack scenario}}} & \emph{\textbf{HW}} & \emph{\textbf{SB}} & \emph{\textbf{NT}} & \emph{\textbf{SNA}} & \emph{\textbf{RSR}} & \emph{\textbf{FNC}} & \emph{\textbf{INT}} & \emph{\textbf{INB}} & \emph{\textbf{OUT}} & \emph{\textbf{SRD}} \\
\hhline{===========}
C\&C communi.
& 40\% & 60\% & \textbf{90\%} & 10\% & 20\% & 30\% & 20\% & 70\% & \textbf{90\%} & \textbf{80\%} \\ \hline
DDoS flooding & 54\% & 62\% & \textbf{100\%} & 15\% & 31\% & 46\% & 38\% & 46\% & \textbf{85\%} & \textbf{85\%} \\ \hline
Data exfiltration & 50\% & 50\% & \textbf{100\%} & 33\% & 33\% & 25\% & 25\% & 0\% & 75\% & \textbf{83\%} \\ \hline
Bot scanning & 40\% & \textbf{80\%} & \textbf{80\%} & 0\% & 40\% & \textbf{80\%} & 20\% & 60\% & \textbf{80\%} & \textbf{80\%} \\ 
\hhline{===========}
\multicolumn{1}{c|}{\textbf{Average}} & 46\% & 63\% & \textbf{93\%} & 15\% & 31\% & 45\% & 26\% & 44\% & \textbf{82\%} & \textbf{82\%} \\ \hhline{===========}
\end{tabular}%
}
\end{table}

\begin{figure}[!h]
\centerline{\includegraphics[width=0.9\linewidth]{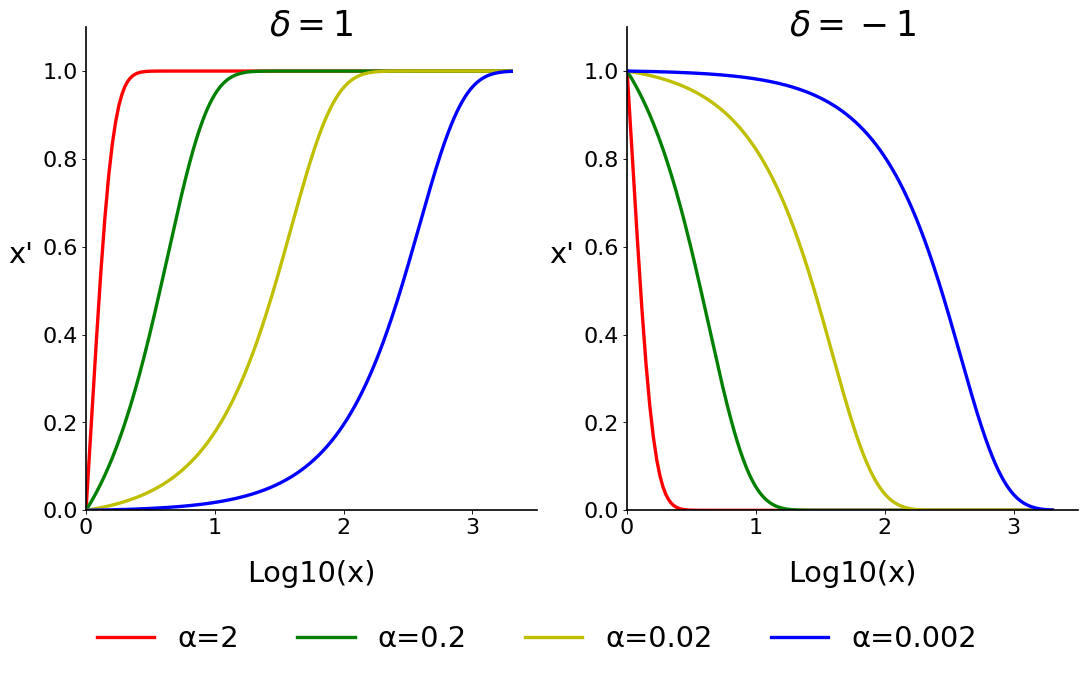}}
\caption{Feature normalization for various ranges.}\label{fig:tanh}
\end{figure}

As shown in Fig.~\ref{fig:tanh}, all of the $x'$s converge to one, each with a proper steepness based on its $\alpha$. 
That is, the convergence is quicker for $x_i$s whose $x_i^{max}$ is lower.

\subsection{Computing the Detectability Score of a Given IoT Model for an Attack Scenario}\label{subsec:Computing_the_Detectability_Score_of_a_Given_IoT_Device_for_an_Attack_Scenario}

Given combinations of \emph{(a)} the characteristics of an IoT model (as specified in the taxonomy), and \emph{(b)} an attack scenario of interest, this step quantitatively assesses the ease of attack detection. This step's input (i.e., the required technical and behavioral information), can be obtained from the device’s spec sheet, the FCC ID database~\cite{fcc_id}, MUD files~\cite{mud2019}, and from locally captured network traffic. Then, for an attack scenario $A_j$ and IoT model $M$ with features $X'_j$  (normalized for $A_j$), the D-Score is  calculated using Eq.~\eqref{eq:d_score_calculation}.

\begin{equation}
D-Score(A_j, M) = W_j \cdot X'_j = \sum_{i=1}^{|X|} (w_{ij} \cdot x'_{ij})
 \label{eq:d_score_calculation}
\end{equation}

\section{Quantitative Evaluation}\label{sec:evaluation_method_and_validation}

\subsection{Implementation of the Questionnaire}\label{subsec:questionnaire}

In order to quantitatively evaluate our method for D-Score calculation, we designed and implemented an online questionnaire~\cite{questionnaire} using the open-source survey tool LimeSurvey~\cite{schmitz_2017}. The first part of the questionnaire presented some background information and guidelines to the respondent, and in the second part, the respondent was asked to anonymously provide demographic details regarding their professional experience and educational level. In the third part, in order to strengthen the validity of the results, we asked the respondent to select an IoT attack scenario from the list of attacks presented in the threat model (see Table~\ref{tab:attack_scenarios}), based on his/her knowledge.

In the fourth part of the questionnaire, as part of the novel process of preliminary filtering (described in Section~\ref{subsec:deriving_a_detectability_model_for_an_iot_attack_scenario}), the respondent was asked to \emph{(a)} select only the categories relevant for the detection of the attack scenario he/she selected and then to \emph{(b)} select only the relevant sub-categories within the categories selected in \emph{(a)}. 
As summarized in Table~\ref{tab:preliminary_filtering_percentage}, this ability to select specific relevant categories and sub-categories was widely exercised by the respondents. For example, of those who addressed the \emph{Data exfiltration} scenario, only 50\% kept the \emph{SB} category, (i.e. the remaining 50\% filtered it out), leading to a reduction of 11 AHP-style pairwise comparisons among the sub-categories, three comparisons among the features in the \emph{FNC} sub-category and 10 more among the features in the \emph{INT} sub-category, i.e., a total reduction of 24 (!) pairwise comparisons. Moreover, even if a category was not filtered out, often the respondent did not select all of the sub-categories in the category. For example, for \emph{Data exfiltration} 100\% of the respondents kept the \emph{NT} category, however 0\% kept the \emph{INB} sub-category, thus reduced their total comparisons by 12.

In the fifth part of the questionnaire each respondent was asked to perform pairwise comparisons among the remaining sub-categories as well as among the features within these sub-categories. The scale of each comparison ranges from -5 to +5, where ``5'' signifies extreme importance, ``3'' reflects essential or strong importance, and ``0'' means equal importance. The ``-'' and ``+'' signs determine which is more important, i.e., the left-hand side or the right-hand side of the comparison, respectively.

\subsection{Collection and Preprocessing of the Responses}\label{subsec:response_collection}

To collect responses, we reached out via email to cyber-security experts from our academic and professional circles in Europe, the Middle East, and the Far East. Each expert received a link to our questionnaire~\cite{questionnaire}, accompanied by concise instructions.
After the experts completed the online questionnaires, we exported their responses to a central .csv file. 
Due to the preliminary filtering process used in the fourth part, there were missing values in this .csv file for the categories and sub-categories filtered out by the experts. 
Therefore, as part of our data preprocessing, we filled in the missing values as follows: If a category or sub-category was kept but not the other one, a value of $\pm 5$ was added, depending on the side of the pairwise comparison. If neither category or sub-category was kept, then a value of 0 was added, meaning they were equally unimportant.

\begin{figure}[t]
\centerline{\includegraphics[width=1.0\linewidth, trim={0.1cm 0.1cm 0.1cm 0.1cm},clip]{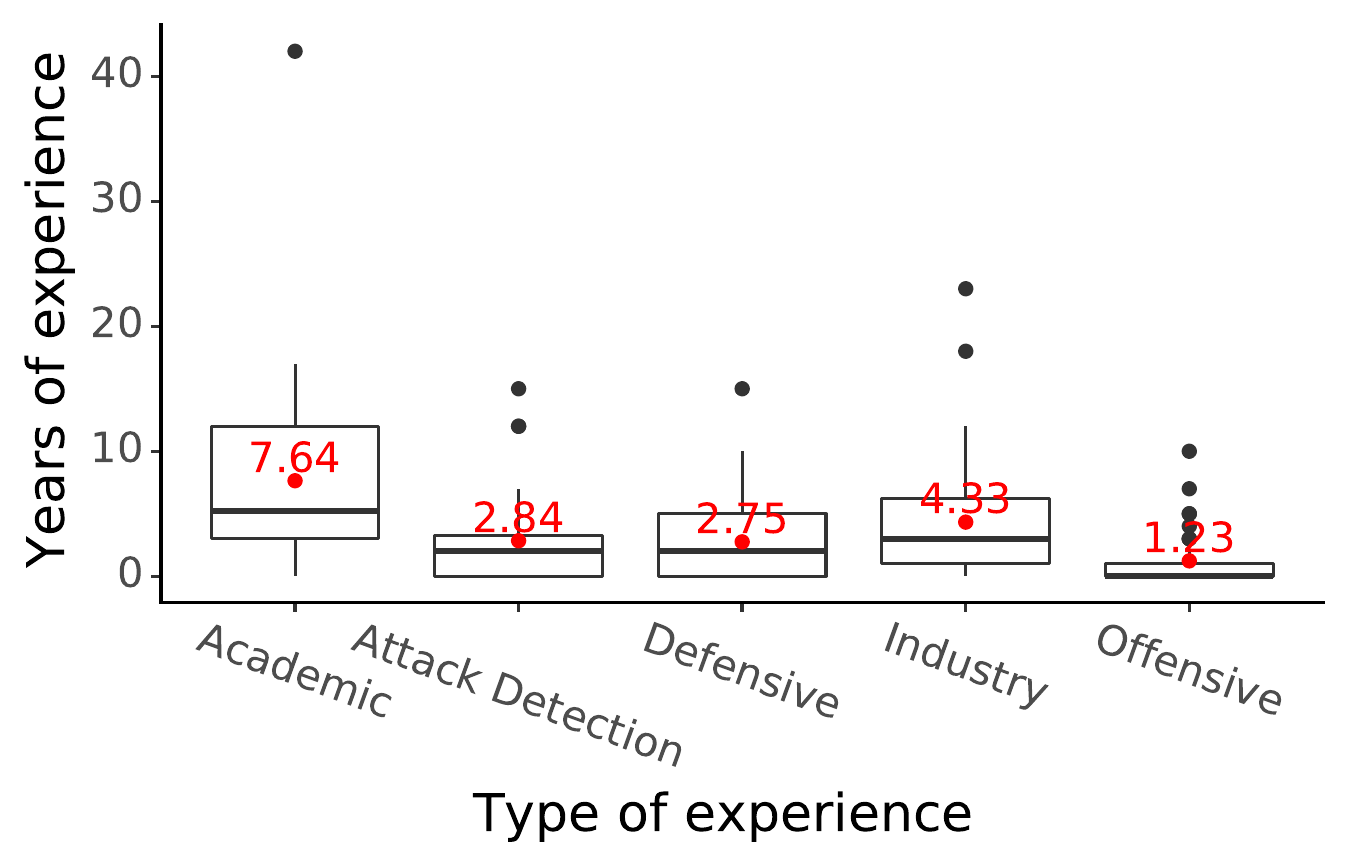}}
\caption{Background of the cyber-security experts.}
\label{fig:experts_experience}
\end{figure}

\begin{figure*}[ht]
\centerline{\includegraphics[width=0.8\textwidth, trim={3cm 8cm 3cm 2.5cm},clip]{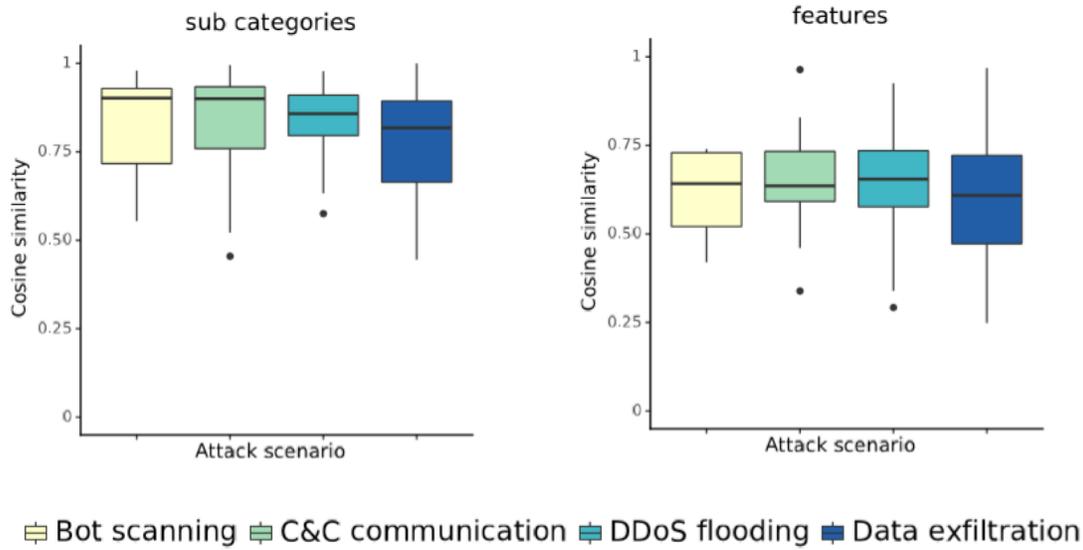}}
\caption{Levels of agreement among the cyber-security experts for the evaluated attack scenarios.}
\label{fig:agreement_sub_categories}
\end{figure*}

\begin{figure*}[!b]
\centerline{\includegraphics[width=1.0\textwidth, trim={0 0.3cm 0 0}, clip]{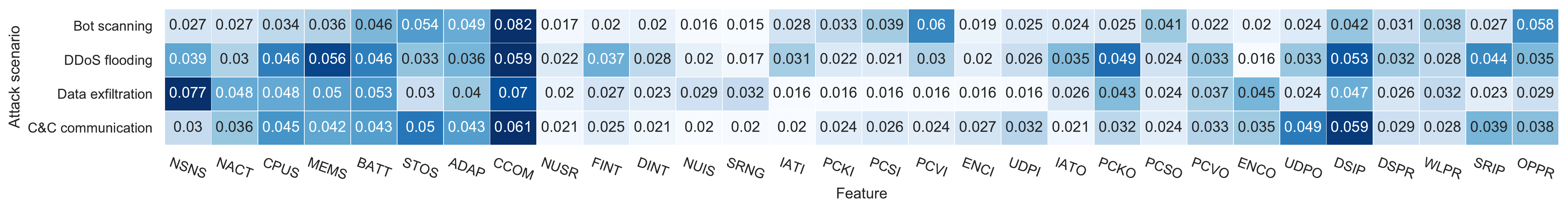}}
\caption{Weights of the features across the evaluated attack scenarios}\label{fig:feature_weights}
\end{figure*}

The median amount of time it took for a respondent to complete the questionnaire was only 13 minutes. Based on the collected demographic details, in Fig.~\ref{fig:experts_experience} we can see that our respondents have more academic experience (7.64 years, on average) than industry experience (4.33), and more experience in defensive cyber-security (2.75 years, on average) than offensive cyber-security (1.23). In addition, of the 40 respondents, 18 have a doctoral degree, and 22 have a master's degree (many of whom are currently enrolled in Ph.D. programs) or less (mostly M.Sc. or B.Sc. students at the time of response collection).

\subsection{Calculation of the Weights Using AHP}\label{subsec:calc_weights_using_ahp}

To analyze the data using the AHP methodology, we implemented a Python script which is capable of receiving the pairwise comparisons of a respondent as input and producing the weights of the features and sub-categories as output. We applied this script to each response separately and then calculated the average weight for each of the four groups of responses corresponding to the four attack scenarios.

As noted above, we collected multiple responses from various cyber-security experts for each of the four IoT attack scenarios. As an indication of the responses' validity and robustness, for each attack scenario we measured the \emph{agreement level} between each pair of weight vectors $W$ in terms of the average cosine similarity, defined in Eq.~\eqref{eq:cosine_similarity}, similarly to~\cite{bitton2018taxonomy}.

\begin{equation}
Cosine \,  Similarity(A, B) = \frac{A \cdot B}{\left \| A \right \|\left \| B \right \|} \label{eq:cosine_similarity}
\end{equation}

As evident in Fig.~\ref{fig:agreement_sub_categories}, the agreement level as to the relative importance of the sub-categories is very high. It is also evident that the agreement level for the \emph{Data exfiltration} scenario is moderately lower than the other three, maybe due to a lower level of attack scenario clarity. That is, \emph{C\&C communication}, \emph{DDoS flooding} and \emph{Bot scanning} may all be very unambiguous and well-known to cyber-security experts. 
In contrast, there is a chance that the respondents interpreted the \emph{Data exfiltration} scenario in different ways, thus ranked the related sub-categories differently from one another.
Regarding the weighing of features, it is apparent how the agreement levels for all of the attack scenarios are shifted downwards, compared to the sub-categories. The reason may be that pairwise comparisons among a large number of distinct elements (30 features) may be more complicated than comparing seven sub-categories to one another.

The feature weights $W_j$ for each evaluated IoT attack scenario $A_j$ are presented in Fig.~\ref{fig:feature_weights}. Each weight is an average of the corresponding weights given by all of the experts. In this heat map, the color is normalized for every row separately so that for every attack scenario we can easily observe the features that are found most informative by the experts. On the one hand, features like \emph{CCOM} seem very informative for all of the attack scenarios evaluated, while on the other hand, features like \emph{NUSR} or \emph{ENCI} may be of negligible importance for detecting any of the attack scenarios evaluated. 
Though uninformative for these scenarios, we advise against removing them completely from the taxonomy, since they may be relevant for other attacks to be addressed in the future. In between, there are features that are deemed important for some attacks and not for others. For instance, \emph{PCVI} has the largest weights for \emph{bot scanning} detection and relatively low weights for the other scenarios, likely because this is where anomalies would be found once this attack scenario is realized. Interestingly, for all of the evaluated attacks, the indirect features under the \emph{Hardware} category (i.e., NSNS, NACT, CPUS, MEMS, and BATT) were assigned relatively high $W$s, meaning that this novel part of our taxonomy was found to be informative by the experts for IoT attack detection, even though they are not direct/detectable attack features~\cite{kovanen2016survey}.

\begin{table*}[!b]
\centering
\caption{D-Scores of alternative security cameras studied in~\cite{meidan2018nbaiot} in the context of a data-driven anomaly-based method to detect DDoS flooding attacks executed by IoT devices.}
\label{tab:validation_nbaiot}
\resizebox{\textwidth}{!}{%
\begin{tabular}{l|l|l|r|c|r|r|rrrrr}
\hhline{============}
\multicolumn{3}{c|}{\textit{\textbf{Evaluated IoT models}}} &
  \multicolumn{2}{c|}{\textit{\textbf{Detectability}}} &
  \multicolumn{3}{c|}{\textit{\textbf{Algorithms' window sizes}}} &
  \multicolumn{4}{c}{\textit{\textbf{Key differentiating features}}} \\ \hline
\multicolumn{1}{c|}{\textit{\textbf{Ref.}}} &
  \multicolumn{1}{c|}{\textit{\textbf{Manufacturer}}} &
  \multicolumn{1}{c|}{\textit{\textbf{Model num.}}} &
  \multicolumn{1}{c|}{\textit{\textbf{Score}}} &
  \textit{\textbf{Label}} &
  \multicolumn{1}{c|}{\textit{\textbf{IF}}} &
  \multicolumn{1}{c|}{\textit{\textbf{LOF}}} &
  \multicolumn{1}{c|}{\textit{\textbf{OCSVM}}} &
  \multicolumn{1}{c|}{\textit{\textbf{CCOM}}} &
  \multicolumn{1}{c|}{\textit{\textbf{IATO}}} &
  \multicolumn{1}{c|}{\textit{\textbf{PCVO}}} &
  \multicolumn{1}{c}{\textit{\textbf{DSPR}}} \\ \hhline{============}
~\cite{PT838E} &
  Provision &
  PT-838 &
  0.429 &
  E &
  170 &
  1,300 &
  \multicolumn{1}{r|}{350} &
  \multicolumn{1}{r|}{0.900} &
  \multicolumn{1}{r|}{8.655} &
  \multicolumn{1}{r|}{142.033} &
  12 \\ \hline
~\cite{SimpleHome1003} &
  Simple Home &
  XCS7-1003 &
  0.433 &
  E &
  170 &
  600 &
  \multicolumn{1}{r|}{250} &
  \multicolumn{1}{r|}{0.820} &
  \multicolumn{1}{r|}{25.035} &
  \multicolumn{1}{r|}{131.895} &
  10 \\ \hline
~\cite{PT737E} &
  Provision &
  PT-737E &
  0.462 &
  \textbf{D} &
  96 &
  \textbf{200} &
  \multicolumn{1}{r|}{91} &
  \multicolumn{1}{r|}{0.960} &
  \multicolumn{1}{r|}{\textbf{15.673}} &
  \multicolumn{1}{r|}{\textbf{122.780}} &
  \textbf{9} \\ \hline
~\cite{SimpleHome1002} &
  Simple Home &
  XCS7-1002 &
  \textbf{0.477} &
  \textbf{D} &
  \textbf{61} &
  300 &
  \multicolumn{1}{r|}{\textbf{86}} &
  \multicolumn{1}{r|}{\textbf{0.370}} &
  \multicolumn{1}{r|}{25.060} &
  \multicolumn{1}{r|}{125.956} &
  40 \\ \hhline{============}
\multicolumn{5}{r|}{\textit{\textbf{Correlation with the D-Score}}} &
  -0.997 &
  -0.809 &
  -0.943 &
   &
   &
   &
   \\ \hhline{============}
\end{tabular}%
}
\end{table*}

\begin{figure}[t]
\centerline{\includegraphics[width=1.0\linewidth, trim={0 0.2cm 0 0}, clip]{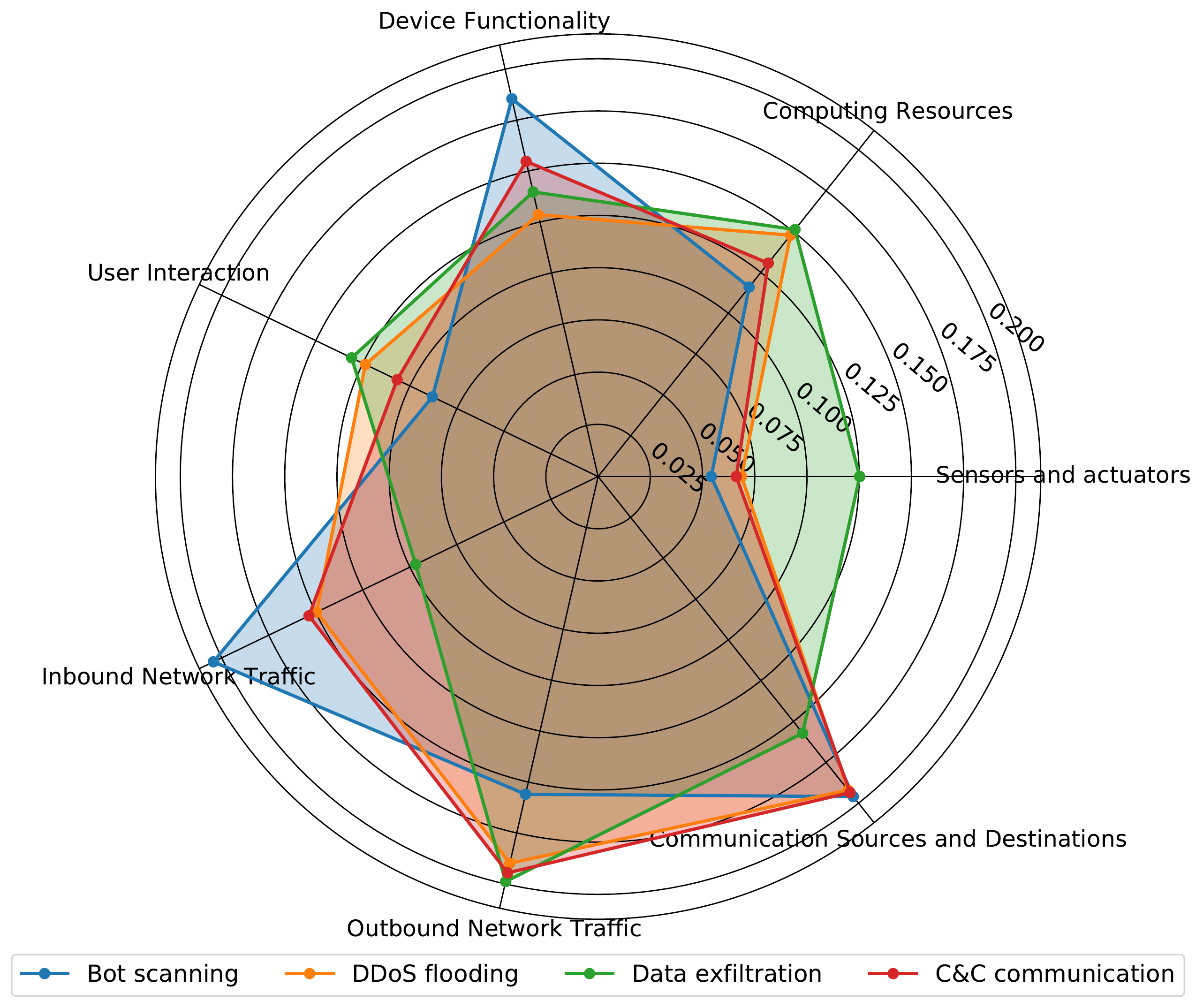}}
\caption{Weights of the sub-categories across the evaluated attack scenarios.}
\label{fig:subcategories_radar}
\end{figure}

Fig.~\ref{fig:subcategories_radar} presents the weights calculated for the sub-categories. As expected, the sub-categories which contain the direct/detectable features (i.e., the \emph{Inbound} and \emph{Outbound network traffic} as well as the \emph{Communication sources and destinations}) were ranked as more informative for attack detection than the indirect features. Among them, the \emph{Inbound network traffic} was given the highest weight for \emph{Botnet scanning}, likely due to the nature of this scenario, in which bots try to communicate with (i.e., scan) the IoT device. Nevertheless, the sub-categories which involve the indirect features were assigned non-negligible weights. This is most apparent for the \emph{Sensors and Actuators} sub-category, which was found to be informative for detecting \emph{Data exfiltration} attacks. This is probably because higher values of the \emph{Number of sensors} feature lead to more frequent and versatile outbound traffic (thus less predictable), which could hide this attack.

\begin{table*}[ht]
\centering
\caption{DDoS flooding attack detection: Correlation between expert-based D-Scores of various IoT device types and models, and data-driven anomaly-based performance metrics obtained in~\cite{meidan2018nbaiot}.}
\label{tab:validation_nbaiot_multiple_types}
\resizebox{0.8\textwidth}{!}{%
\begin{tabular}{clllrc|rrr}
\hhline{=========}
\multicolumn{4}{c|}{\textit{\textbf{Evaluated IoT models}}} &
  \multicolumn{2}{c|}{\textit{\textbf{Detectability}}} &
  \multicolumn{3}{c}{\textit{\textbf{Algorithms' window sizes}}} \\ \hline
\multicolumn{1}{c|}{\textit{\textbf{Ref.}}} &
  \multicolumn{1}{c|}{\textit{\textbf{Type}}} &
  \multicolumn{1}{c|}{\textit{\textbf{Manufacturer}}} &
  \multicolumn{1}{c|}{\textit{\textbf{Model num.}}} &
  \multicolumn{1}{c|}{\textit{\textbf{Score}}} &
  \textit{\textbf{Label}} &
  \multicolumn{1}{c|}{\textit{\textbf{IF}}} &
  \multicolumn{1}{c|}{\textit{\textbf{LOF}}} &
  \multicolumn{1}{c}{\textit{\textbf{OCSVM}}} \\ \hhline{=========}
\multicolumn{1}{c|}{~\cite{philipsb120}} &
  \multicolumn{1}{l|}{Baby monitor} &
  \multicolumn{1}{l|}{Philips} &
  \multicolumn{1}{l|}{B120N/10} &
  \multicolumn{1}{r|}{0.424} &
  E &
  \multicolumn{1}{r|}{170} &
  \multicolumn{1}{r|}{1,800} &
  300 \\ \hline
\multicolumn{1}{c|}{~\cite{PT838E}} &
  \multicolumn{1}{l|}{Camera} &
  \multicolumn{1}{l|}{Provision} &
  \multicolumn{1}{l|}{PT-838} &
  \multicolumn{1}{r|}{0.429} &
  E &
  \multicolumn{1}{r|}{170} &
  \multicolumn{1}{r|}{1,300} &
  350 \\ \hline
\multicolumn{1}{c|}{~\cite{SimpleHome1003}} &
  \multicolumn{1}{l|}{Camera} &
  \multicolumn{1}{l|}{Simple Home} &
  \multicolumn{1}{l|}{XCS7-1003} &
  \multicolumn{1}{r|}{0.433} &
  D &
  \multicolumn{1}{r|}{170} &
  \multicolumn{1}{r|}{600} &
  250 \\ \hline
\multicolumn{1}{c|}{~\cite{danmini}} &
  \multicolumn{1}{l|}{Doorbell} &
  \multicolumn{1}{l|}{Danmini} &
  \multicolumn{1}{l|}{WF 720P} &
  \multicolumn{1}{r|}{0.445} &
  D &
  \multicolumn{1}{r|}{16} &
  \multicolumn{1}{r|}{300} &
  141 \\ \hline
\multicolumn{1}{c|}{~\cite{PT737E}} &
  \multicolumn{1}{l|}{Camera} &
  \multicolumn{1}{l|}{Provision} &
  \multicolumn{1}{l|}{PT-737E} &
  \multicolumn{1}{r|}{0.462} &
  D &
  \multicolumn{1}{r|}{96} &
  \multicolumn{1}{r|}{200} &
  91 \\ \hline
\multicolumn{1}{c|}{~\cite{ennio}} &
  \multicolumn{1}{l|}{Doorbell} &
  \multicolumn{1}{l|}{Ennio} &
  \multicolumn{1}{l|}{Bell} &
  \multicolumn{1}{r|}{0.470} &
  D &
  \multicolumn{1}{r|}{11} &
  \multicolumn{1}{r|}{450} &
  36 \\ \hline
\multicolumn{1}{c|}{~\cite{SimpleHome1002}} &
  \multicolumn{1}{l|}{Camera} &
  \multicolumn{1}{l|}{Simple Home} &
  \multicolumn{1}{l|}{XCS7-1002} &
  \multicolumn{1}{r|}{0.477} &
  D &
  \multicolumn{1}{r|}{61} &
  \multicolumn{1}{r|}{300} &
  86 \\ \hhline{=========}
\multicolumn{6}{r|}{\textit{\textbf{Correlation with the D-Score}}} &
  \multicolumn{1}{r|}{-0.762} &
  \multicolumn{1}{r|}{-0.771} &
  -0.930 \\ \hhline{=========}
\end{tabular}%
}
\end{table*}

\begin{table*}[!b]
\centering
\caption{Calculation of D-Scores for IoT models studied in~\cite{meidan2018nbaiot} and the IoT attack scenarios outlined in Table~\ref{tab:attack_scenarios}.}
\label{tab:d_scores_all_attacks_all_devices}
\resizebox{\textwidth}{!}{%
\begin{tabular}{llll||rrrrrr}
\hhline{==========}
\multicolumn{4}{c||}{\textit{\textbf{IoT model}}} &
  \multicolumn{6}{c}{\textit{\textbf{D-Score}}} \\ \hline
\multicolumn{1}{c|}{\textit{\textbf{Type}}} &
  \multicolumn{1}{c|}{\textit{\textbf{Ref.}}} &
  \multicolumn{1}{c|}{\textit{\textbf{Manufacturer}}} &
  \multicolumn{1}{c||}{\textit{\textbf{Model num.}}} &
  \multicolumn{1}{c|}{\textit{\textbf{DDoS flooding}}} &
  \multicolumn{1}{c|}{\textit{\textbf{Bot scanning}}} &
  \multicolumn{1}{c|}{\textit{\textbf{Data exfiltration}}} &
  \multicolumn{1}{c||}{\textit{\textbf{C\&C communication}}} &
  \multicolumn{1}{c|}{\textit{\textbf{Avg.}}} &
  \multicolumn{1}{c}{\textit{\textbf{(Maxi-)min.}}} \\ \hhline{==========}
\multicolumn{1}{l|}{\multirow{4}{*}{Camera}} &
  \multicolumn{1}{l|}{~\cite{PT737E}} &
  \multicolumn{1}{l|}{Provision} &
  PT-737E &
  \multicolumn{1}{r|}{0.462} &
  \multicolumn{1}{r|}{0.426} &
  \multicolumn{1}{r|}{0.493} &
  \multicolumn{1}{r||}{0.470} &
  \multicolumn{1}{r|}{0.463} &
  0.426 \\ \cline{2-10} 
\multicolumn{1}{l|}{} &
  \multicolumn{1}{l|}{~\cite{PT838E}} &
  \multicolumn{1}{l|}{Provision} &
  PT-838 &
  \multicolumn{1}{r|}{0.429} &
  \multicolumn{1}{r|}{0.408} &
  \multicolumn{1}{r|}{0.478} &
  \multicolumn{1}{r||}{0.446} &
  \multicolumn{1}{r|}{0.440} &
  0.408 \\ \cline{2-10} 
\multicolumn{1}{l|}{} &
  \multicolumn{1}{l|}{~\cite{SimpleHome1002}} &
  \multicolumn{1}{l|}{Simple Home} &
  XCS7-1002 &
  \multicolumn{1}{r|}{\textbf{0.477}} &
  \multicolumn{1}{r|}{\textbf{0.460}} &
  \multicolumn{1}{r|}{\textbf{0.502}} &
  \multicolumn{1}{r||}{\textbf{0.489}} &
  \multicolumn{1}{r|}{\textbf{0.482}} &
  \textbf{0.460} \\ \cline{2-10} 
\multicolumn{1}{l|}{} &
  \multicolumn{1}{l|}{~\cite{SimpleHome1003}} &
  \multicolumn{1}{l|}{Simple Home} &
  XCS7-1003 &
  \multicolumn{1}{r|}{0.433} &
  \multicolumn{1}{r|}{0.418} &
  \multicolumn{1}{r|}{0.466} &
  \multicolumn{1}{r||}{0.461} &
  \multicolumn{1}{r|}{0.445} &
  0.418 \\ \hhline{==========}
\multicolumn{1}{l|}{\multirow{2}{*}{Doorbell}} &
  \multicolumn{1}{l|}{~\cite{danmini}} &
  \multicolumn{1}{l|}{Danmini} &
  WF 720P &
  \multicolumn{1}{r|}{0.445} &
  \multicolumn{1}{r|}{0.418} &
  \multicolumn{1}{r|}{0.460} &
  \multicolumn{1}{r||}{0.449} &
  \multicolumn{1}{r|}{0.443} &
  0.418 \\ \cline{2-10} 
\multicolumn{1}{l|}{} &
  \multicolumn{1}{l|}{~\cite{ennio}} &
  \multicolumn{1}{l|}{Ennio} &
  Bell &
  \multicolumn{1}{r|}{\textbf{0.470}} &
  \multicolumn{1}{r|}{\textbf{0.429}} &
  \multicolumn{1}{r|}{\textbf{0.498}} &
  \multicolumn{1}{r||}{\textbf{0.480}} &
  \multicolumn{1}{r|}{\textbf{0.469}} &
  \textbf{0.429} \\ \hhline{==========}
\multicolumn{1}{l|}{Baby monitor} &
  \multicolumn{1}{l|}{~\cite{philipsb120}} &
  \multicolumn{1}{l|}{Philips} &
  B120N/10 &
  \multicolumn{1}{r|}{\textbf{0.424}} &
  \multicolumn{1}{r|}{\textbf{0.403}} &
  \multicolumn{1}{r|}{\textbf{0.437}} &
  \multicolumn{1}{r||}{\textbf{0.398}} &
  \multicolumn{1}{r|}{\textbf{0.416}} &
  \textbf{0.398} \\ \hhline{==========}
\end{tabular}%
}
\end{table*}

\subsection{Utilizing the Weights to Calculate D-Scores}\label{subsec:Utilizing_the_Weights_to_Calculate_D_Scores}

Having calculated the weights $W$ using AHP, in this subsection we demonstrate how an organization can utilize them to calculate D-Scores for alternative, competing, IoT models and decide accordingly which to deploy. For that, we purchased four security cameras whose models were studied in~\cite{meidan2018nbaiot}. We deployed them in our lab and collected their static and dynamic values which correspond to the 30 features in the taxonomy. Assuming that \emph{DDoS flooding} is the primary attack scenario to defend from, we calculated the respective D-Scores. Table~\ref{tab:validation_nbaiot} presents the resultant D-Scores and also compares them to a key attack detection metric obtained from~\cite{meidan2018nbaiot}. In that paper, several anomaly detection algorithms were used to detect flooding attacks carried out by Mirai and BASHLITE botnets from commercial IoT devices, including Isolation Forest~\cite{liu2008isolation} (abbreviated as IF in Table~\ref{tab:validation_nbaiot}), Local Outlier Factor~\cite{breunig2000lof} (LOF) and One-Class Support Vector Machine (OCSVM)~\cite{noumir2012simple}. The anomaly decisions were taken in that paper based on a \emph{sequence} of traffic metadata feature vectors, using a majority vote on a moving window whose length was optimized for each IoT model separately. The authors of~\cite{meidan2018nbaiot} demonstrated excellent detection results in terms of true positive rate (TPR)
and false positive rate (FPR), however with a cost of relatively 
lengthy moving windows (equivalent to long detection times), treated here as a metric for attack detectability.

As evident in Table~\ref{tab:validation_nbaiot}, for the DDoS flooding attack we found a strong negative correlation between the (optimized) window sizes and the D-Scores for each evaluated algorithm. This means that, as expected, D-Scores which we attained using an expert-based method are aligned with an attack detectability metric obtained using a data-driven method, such that longer window sizes (i.e., extended detection times) are expressed via lower D-Scores. In addition, among the four security camera models in Table~\ref{tab:validation_nbaiot}, the D-Score of~\cite{PT838E} was found to be the lowest, meaning that worse detection performance is expected from this IoT model for DDoS flooding attacks, compared with the other models. This finding corresponds well with~\cite{meidan2018nbaiot}, where the optimized window size is the highest for this model for all the evaluated algorithms.

As described in Section~\ref{sec:introduction}, a (continuous) D-Score can be translated into a (categorical) detectability label. If we employ the same number of categories as in Fig.~\ref{subfig:label_example} (inspired by~\cite{wikityre}), and use equal-width binning for discretization, then seven bins are attained, each is $1/7=0.143$ wide. In that case, an IoT model whose $D-Score\in[0.857, 1.00]$ would be labeled as ``A'' (most highly detectable, thus advised for deployment). Regarding the IoT models in our evaluation (i.e., the four security cameras), Table~\ref{tab:validation_nbaiot} indicates that two are labaled as ``E'' since their $D-Score\in[0.286, 0.429]$, and the other two as ``D'' (superior detectability of DDoS flooding attacks, $D-Score\in[0.429, 0.571]$).

Table~\ref{tab:validation_nbaiot} also provides key features whose values differentiate among the four alternative security cameras. For instance, when comparing the security camera with the lowest D-Score~\cite{PT838E} to the one with the highest~\cite{SimpleHome1002}, it is evident how in most cases lower D-Scores are associated with network traffic that is more frequent and variable, e.g., a higher tendency to communicate continuously (\emph{CCOM}), shorter inter-arrival times of outbound sessions (\emph{IATO}), higher variability of outbound packet size (\emph{PCVO}), and higher number of destination ports (\emph{DSPR}). Another interesting finding is that among the 30 taxonomy features, the ones which differentiate the most between the normal behavior of the above two camera models also conform with past research. For instance, the top three features in terms of absolute difference in $x'$ (namely IATI, PCKI and PCVI) have been found informative by~\cite{shaikh2019iot} for IoT attack detection. Similarly, IATI and PCVI have been selected by~\cite{doshi2018machine} as two of the three most important features, using Gini Score.

\subsection{Extension of the DDoS Flooding Attack Detectability Evaluation to Non-Competing IoT Models}\label{subsec:extension_of_the_DDoS_flooding_attack_detectability_evaluation_to_non_competing_IoT_models}

In subsection~\ref{subsec:Utilizing_the_Weights_to_Calculate_D_Scores} we focused on a use-case in which D-Scores are calculated for alternative (competing) IoT models of the same type, to prioritize and determine: which smart camera to deploy, assuming that the organization aims at quick and accurate detection of DDoS flooding attacks. In the current subsection, summarized in Table~\ref{tab:validation_nbaiot_multiple_types}, we extend our evaluation to another use-case, in which the D-Score is used for black/white listing of non-competing IoT devices. That is, an organization's security policy can set a threshold on the D-Score, and only the IoT devices whose D-Scores surpass this threshold are allowed to be connected to the organization network. As evident in Table~\ref{tab:validation_nbaiot_multiple_types}, which adds a smart baby monitor and two doorbells to the four previously-analyzed cameras, strong negative correlations are maintained between our expert-based D-Scores and the data-driven performance metric from N-BaIoT~\cite{meidan2018nbaiot}. Despite the larger variability in IoT device types and models, lower D-Scores are still associated with longer window sizes (detection times), most strongly for the OCSVM-based AIDS.

\subsection{Extension of the Detectability Evaluation to Additional IoT Attack Scenarios}\label{subsec:Extension_of_the_detectability_evaluation_to_additional_IoT_attack_scenarios}

Until now, in Subsections~\ref{subsec:Utilizing_the_Weights_to_Calculate_D_Scores} and~\ref{subsec:extension_of_the_DDoS_flooding_attack_detectability_evaluation_to_non_competing_IoT_models}, our quantitative evaluation concentrated only on one of the four attack scenarios we address in this paper (outlined in Table~\ref{tab:attack_scenarios}), namely DDoS flooding attack. The reason is that this is the only attack scenario for which we have data-driven performance metrics~\cite{meidan2018nbaiot} to compare with our expert-based approach. In this subsection, for completeness, we calculate the D-Score for the same IoT models evaluated in Subsection~\ref{subsec:extension_of_the_DDoS_flooding_attack_detectability_evaluation_to_non_competing_IoT_models}, this time for the remaining attack scenarios, namely bot scanning, data exfiltration and C\&C communication. The summary of this evaluation is portrayed in Table~\ref{tab:d_scores_all_attacks_all_devices}, where for each IoT type (camera, doorbell and baby monitor) we highlighted \emph{(a)} the highest D-Score for each IoT attack scenario, \emph{(b)} the highest average D-Score, and \emph{(c)} the highest minimum D-Score. The latter is highlighted in order to recognize the IoT model whose minimum D-Score for any attack scenario of interest is higher than the other (competing) IoT models. In real-world settings, this \emph{maximin criterion}~\cite{barbara1988maximin} could support procurement and deployment decisions, based on maximizing the minimum chance of detecting an attack carried out.

The results (see Table~\ref{tab:d_scores_all_attacks_all_devices}) show that when contemplating which smart camera to deploy, Simple Home XCS7-1002 consistently has the maximum chances for attack detection, and so is Ennio Bell (for smart doorbells) and the Philips B120N/10 baby monitor. This may be due to the fact that Simple Home XCS7-1002 is the least continuously-communicating camera (CCOM), and also has the largest inter-arrival time of outbound sessions (IATI). Similarly, Ennio Bell has the lowest CCOM and IATI, as well as the smallest number of unique destination ports (DSPR). Overall, the maximin D-Score of Philips B120N/10 was the lowest of all IoT models evaluated, meaning that any of the attack scenarios addressed in this paper would be the hardest to detect. We attribute this low attack detectability score mostly to the largest percentage of outbound encrypted traffic (ENCO) and to the highest number of sensors (NSNS), which include video, motion detection, noise detection, temperature sensor and humidity sensor.

Regarding the findings in this subsection, we note that except for the DDoS flooding attack, the D-Score results for the remaining attack scenarios are expert-based estimations and need to be validated using a data driven approach, once proper data and labels are available. We leave this for future research.

\subsection{Analysis of the Sensitivity to the Quality of Questionnaire Responses}\label{subsec:Model_Tuning_and_Sensitivity_Analysis}

The experimental results discussed thus far were attained based on \emph{all} of the questionnaire responses we received, as reported in Table~\ref{tab:attack_scenarios}. However, not all of the respondents have the same background, experience and level of expertise in cyber-security, IoT or attack detection. Therefore, a question arises regarding the proper selection of responses, such that safe feature weights are eventually drawn, to be used for D-Score calculation. To address this question, in this subsection we analyze the sensitivity of our method to using only \emph{subsets} of the responses, based on varying thresholds of response quality criteria. As the range of the explicit criteria such as the years of experience (see Fig.~\ref{fig:experts_experience}) is relatively limited, we chose to use the more implicit response quality criterion known as \emph{consistency ratio} (CR). The CR~\cite{saaty1988analytic} is calculated for a group of pairwise comparisons (e.g., comparisons among multiple features within a sub-category), and it reflects the extent of inconsistent judgments made by evaluators in tasks of multiple pairwise comparisons. Typically, CR levels of up to 0.1 are acceptable~\cite{shih2003method, siboni2020ranking}.

In Table~\ref{tab:validation_nbaiot}, strong negative correlations are demonstrated between the (expert-based) D-Scores of the four security camera models and their (data-driven) optimized window sizes in N-BaIoT~\cite{meidan2018nbaiot}, based on all thirteen questionnaire responses to the \emph{DDoS flooding} attack scenario. To examine the sensitivity of our method to CR, we \emph{(a)} calculated the CR for each of the seven sub-categories in each response, \emph{(b)} calculated the Mean CR for each response, \emph{(c)} calculated D-Scores for varying thresholds of Mean CR, i.e., based on a subset of responses whose Mean CR is below the threshold, and \emph{(d)} calculated the correlation of the D-Scores with the window sizes optimized for each algorithm. The results of this sensitivity analysis are summarized in Table~\ref{tab:CR_thresholding}, and they show some variability of the correlation strength. On the one hand, for the \emph{IF} algorithm the strongest correlation was attained when the D-Scores were calculated using all thirteen responses, including two responses whose Mean CR are 0.172 and 0.198. On the other hand, for the two other algorithms, no more than four (!) responses (the most consistent ones) were required to reach the best results. Looking forward, for feature weights calculation in future applications of the D-Score methodology, remaining with the commonly used threshold of 0.1 seems like a reasonable option. In our case it yields sub-optimal yet strong correlation with the data-driven approach.

\begin{table}[t]
\centering
\caption{Correlation of the D-Scores calculated for the \emph{DDoS flooding} attack scenario with the optimized window sizes attained in N-BaIoT~\cite{meidan2018nbaiot} as a function of the Mean CR threshold used to filter questionnaire responses}
\label{tab:CR_thresholding}
\resizebox{\columnwidth}{!}{%
\begin{tabular}{c|c|c|c|c}
\hhline{=====}
\textit{\textbf{Range}} & \textit{\textbf{Number of}}    & \multicolumn{3}{c}{\textit{\textbf{Correlation of the D-Scores}}}   \\
\textit{\textbf{of}}    & \textit{\textbf{questionnaire}}        & \multicolumn{3}{c}{\textit{\textbf{with the window sizes}}}         \\ \cline{3-5} 
\textit{\textbf{Mean CR}}    & \textit{\textbf{responses}} & \textit{\textbf{IF}} & \textit{\textbf{LOF}} & \textit{\textbf{OCSVM}} \\ 
\hhline{=====}
0 - 0.033 & 4  & -0.981 & \textbf{-0.874} & \textbf{-0.978} \\ \hline
0 - 0.067 & 9  & -0.989 & -0.841 & -0.965 \\ \hline
0 - 0.100 & 11 & -0.994 & -0.829 & -0.956 \\ \hline
All       & 13 & \textbf{-0.997} & -0.809 & -0.943 \\ 
\hhline{=====}
\end{tabular}
}
\end{table}

\begin{table*}[!b]
\centering
\caption{Comparison of the overhead required by the data-driven and the expert-based approaches}
\label{tab:data_driven_vs_expert_based}
\resizebox{\textwidth}{!}{%
\begin{tabular}{l|c|c|l}
\hhline{====}
\multicolumn{1}{c|}{\textit{\textbf{Description of}}} &
  \multicolumn{2}{c|}{\textit{\textbf{Approach}}} &
  \multicolumn{1}{c}{\multirow{2}{*}{\textit{\textbf{Remarks}}}} \\ \cline{2-3}
\multicolumn{1}{c|}{\textit{\textbf{overhead component}}} &
  \textit{\textbf{Data-driven}} &
  \textit{\textbf{Expert-based}} &
  \multicolumn{1}{c}{} \\ \hhline{====}
Procurement of $M$ device(s)                                        & \Checkmark & \Checkmark &                           \\ \hline
Deployment of $M$ device(s) in a lab                                & \Checkmark & \Checkmark &                           \\ \hline
Extraction of static features from, e.g., the spec sheet            &            & \Checkmark &                           \\ \hline
Capturing of benign traffic data                                    & \Checkmark & \Checkmark &                           \\ \hline
Implementation and execution of $A$ attack                          & \Checkmark &            & For model calibration     \\ \hline
Capturing of malicious traffic data                                 & \Checkmark &            & For model calibration     \\ \hline
Precise labeling of traffic data (malicious / benign)               & \Checkmark &            & For model calibration     \\ \hline
Extraction of dynamic features from the captured traffic            & \Checkmark & \Checkmark &                           \\ \hline
\begin{tabular}[c]{@{}l@{}}Procurement and/or installation of an AIDS\\ OR design, training and testing of an anomaly detection algorithm\end{tabular} &
  \Checkmark &
   &
   \\ \hline
Design and implementation of the questionnaire                      &            & \Checkmark & Can use our questionnaire \\ \hline
Collection of questionnaire responses from experts                  &            & \Checkmark &                           \\ \hline
Analysis of the questionnaire responses (i.e., weights calculation) &            & \Checkmark &                           \\ \hhline{====}
\end{tabular}%
}
\end{table*}

\section{Discussion}\label{sec:discussion}

\subsection{IoT Model Complexity, Traffic Predictability and Attack Detectability}\label{subsec:IoT_model_complexity_Traffic_Predictability_and_Attack_Detectability}

The motivation to conduct this research originated from the insight that in many cases, variability exists in the capability of an AIDS to detect the same cyber-attack on differing IoT models. In order to guide the investigation of the source of this variability in IoT attack detection performance, we raised two groups of research questions in Section~\ref{sec:introduction}, namely \emph{(a)} network traffic predictability and \emph{(b)} IoT attack detectability, which complement one another. Our suspicion was that the (indirect) technical characteristics of an IoT Model (e.g., the number of sensors and actuators), as well as its designated functionality and intended modes of user interaction, have an effect on the (direct) network traffic features that are typically analyzed by AIDSs. We argued that IoT models which \emph{(a)} are more technically complex, \emph{(b)} have a more versatile functionality and \emph{(c)} are interacted with more frequently and more diversely, are likely to express less deterministic traffic patterns. Consequently, a traffic-based profile of normal behavior would be less stable, such that the task of anomaly-based attack detection (which is the basic principle in AIDSs) becomes more difficult. In Section~\ref{sec:iot_traffic_predictability} (and Appendix~\ref{apndx:UNSW}) we explored the stability/predictability of IoT traffic patterns and, indeed, found correlations between them and the above-mentioned device complexity factors. For example, the number of sensors is positively correlated with the number of hourly outbound flows, and the CPU speed is correlated with the number of hourly unique destination ports. In Section~\ref{sec:taxonomy} we introduced a novel taxonomy which incorporates the various static and dynamic features of an IoT model. Then, in Sections~\ref{sec:proposed_method} and~\ref{sec:evaluation_method_and_validation}, we presented a method to assess IoT attack detectability, and quantitatively evaluated it, respectively.

Altogether, we found statistical evidence to the link between the complexity of IoT models and their traffic predictability. In addition, our expert-based method for assessing attack detectability, which utilized the above-mentioned IoT model complexity and traffic predictability, demonstrated strong correlation with three anomaly detection algorithms from past (independent, data-driven) research.

\subsection{Trade-Offs in Designing the Taxonomy}\label{subsec:Trade_offs_in_the_Design_of_the_Taxonomy_Features}

Due to the taxonomy design considerations that we outlined in Section~\ref{sec:taxonomy} (e.g., being informative, flexible, self-explanatory, and feasible to deploy), the selection and definition of some features might be sub-optimal for assessing IoT attack detectability. For instance, the \emph{CPU} and \emph{memory utilization per time frame}~\cite{bezerra2019iotds} could be more informative than the actual \emph{CPU speed} or the \emph{memory size} we eventually incorporated into the taxonomy. The reason for this is that in the case of a DDoS flooding attack, the utilization of these key computing resources might (suddenly) increase regardless of the maximum CPU speed or the total memory size. 
However, measuring the CPU and memory utilization for an IoT device (unlike a PC) is exceedingly challenging and might therefore make the calculation of D-Score infeasible. Another example of a potentially more informative feature is the type of sensors (motion detectors, microphones, etc.) as opposed to their quantity, since different sensor types typically have different traffic ratios and contents. The problem is that \emph{sensor type} is a categorical feature, and encoding it into multiple binary features (e.g., \emph{has a motion detector}, \emph{has a microphone}, etc.) would result in a much more complex taxonomy, and would accordingly require many more pairwise comparisons and thus overburden the questionnaire respondents. Yet another feature that could assist in attack detectability quantification is the deployment location of the IoT device, e.g., a normally-crowded room or a rarely-attended warehouse. However, on the one hand this is another categorical feature that would increase the taxonomy's complexity, while on the other hand we already incorporated user interaction features in our taxonomy, so the deployment location itself becomes redundant.

All things considered, the taxonomy we present and use in this paper is the result of several trial-and-error design iterations, conducted in order to fine-tune the taxonomy while balancing the above-mentioned considerations and trade-offs. In future research, additional fine-tuning can be accomplished by re-defining existing features and/or adding possibly-missing ones. In any case, making such feature changes in the taxonomy (and accordingly also in the questionnaire) is an achievable task using the web-based questionnaire we developed.

\subsection{Comparison of Approaches for IoT Attack Detectability Assessment: Expert-Based vs. Data-Driven}\label{subsec:expert_vs_data}

Hypothetically, an organization which already deploys an AIDS and wishes to use the AIDS to quantify the detectability of a certain attack $A$ in advance could opt to employ a data-driven approach. Naturally, this option would allow a straightforward preliminary assessment which is based on the same conditions as in the projected deployment. Nonetheless, despite the anticipated high accuracy, employing a data-driven (anomaly-based) approach comes with a cost of a non-trivial overhead, as elaborated in Table~\ref{tab:data_driven_vs_expert_based}. Among the substantial components of this overhead is the need to acquire a high-quality set of network traffic data which is representative of $M$ as comprehensively as possible. Key steps in acquiring such a dataset are the implementation and execution of $A$, followed by feature extraction and rigorous ground truth labeling of every instance as ``benign'' or ``malicious''. We note that in theory, anomaly-based attack detection methods (as in AIDSs) necessitate only benign instances for training~\cite{garcia2009anomaly}, as they are intended to capture only the normal traffic patterns of $M$, and then label any severe-enough abnormality as an attack. However, in practice, AIDSs tend to suffer from non-negligible FPR~\cite{garcia2009anomaly, 7506774, 8030867, 7878168, li2020enhancing, Bhuyan2017}, meaning that in too many cases, \emph{rare-yet-benign} abnormal events are falsely identified as attacks. Consequently, redundant and potentially harsh countermeasures might be activated in response. To decrease the FPR which is associated with such (unsupervised) anomaly-based methods, one could use labeled data for model calibration or anomaly threshold tuning, as in semi-supervised~\cite{7878168, li2020enhancing} or hybrid~\cite{Bhuyan2017} approaches.

The substantial cost of acquiring labeled data with malicious instances for FPR reduction, can be saved if an organization chooses to use our expert-based method. Additional considerations should also be taken into account while contemplating between data-driven and expert-based approaches:

\begin{itemize}[leftmargin=*]
    \item \emph{Ability to cope with zero-day attacks.} The above-mentioned cost of acquiring labeled data relies on the assumption that the malware of interest is available for local execution in a controlled environment, to be followed by capturing network traffic from $M$ and labeling the extracted feature vectors. However, this is not always the case, as for new variants of malware or attacks, it might take time to capture $A$'s code or replicate it.
    \item \emph{Generalizability.} Even if the malware (whether original or replicated) \emph{is} available for execution, the data-driven approach might over-fit the specific variant of the attack scenario, the (simulated) manner of attack execution, the network structure, the normal user interaction, etc., thus it might fail in generalizing to other environments or conditions. Also, with the data-driven approach, the detectability assessment is based on the specific algorithm(s) used for anomaly detection. In contrast, the feature weights that are trained for the expert-based approach are agnostic of any algorithm.
    \item \emph{Scalability.} Calculating the D-Score of the same $A$ for multiple $M$s using a data-driven method would require to repeat the entire process of device deployment / attack execution / data capturing and labeling / model training and calibration, etc., multiple times, i.e., once for each $M$. In comparison, our method is more efficient in the sense that \emph{(a)} the existing taxonomy and online questionnaire are generic and ready to be used, \emph{(b)} phase A in our method (deriving a detectability model, see Subsection~\ref{subsec:deriving_a_detectability_model_for_an_iot_attack_scenario}) is performed just once for an IoT attack scenario $A$, and \emph{(c)} only phase B (computing the D-Score of an IoT model $M$ for an attack scenario $A$, see Subsection~\ref{subsec:Computing_the_Detectability_Score_of_a_Given_IoT_Device_for_an_Attack_Scenario}) recurs for each $M$ separately. Note that phase B merely requires collecting normal traffic data and some technical characteristics of $M$, and then running an automated script to calculate the final D-Score using AHP. Conversely, calculating D-Score of multiple $A$s for a given $M$ could be instantaneous in cases where phase A (attack-oriented) has already been performed for other $M$s, and phase B (device-oriented) has been performed for the same $M$ in the past, regardless of $A$'s identity. In any case, we assume that any organization which chooses to incorporate D-Scores / labels into its security policies, whether calculated using a data-driven or expert-based method, would focus only on a limited set of $A$s and $M$s, i.e., the top most prevalent or impactful attack scenarios, and a relatively small number of alternative $M$s.
    \item \emph{Objectivity.} Product reviews posted in e-commerce websites could sometimes be manipulated or even fake~\cite{zhuang2018manufactured}. On the one hand, our expert-based method also (partially) relies on human inputs. On the other hand, the participants in the envisioned deployment of our method are independent cyber-security experts, e.g., knowledgeable scholars, experienced practitioners and/or regulatory standardization organizations such as the FCC~\cite{fcc_website}. Thus, a high level of reliability and objectivity is anticipated. 
\end{itemize}

When considering the use of a classic data-driven approach versus our expert-based approach, we make a distinction between two scenarios: In the first scenario, the organization interested in assessing the detectability of IoT attacks has \emph{everything} it needs to perform this task. This includes, e.g., a list of IoT models to consider, a list of IoT-related attacks that are anticipated by and/or dangerous to the organization, enough data collected from the IoT models of interest (which is also accurately labeled with the specific malware variants), an existing attack detection system to be used continuously in research and production, dedicated research staff, the time and funding needed to conduct research while fine-tuning the detection models, and so forth. In this scenario, the organization is more likely to favor the data-driven approach. However, the second scenario may better reflect the reality of many organizations today; in this scenario the organization lacks sufficient IoT security expertise, because, for example, it is a small office or a non-tech company, and is willing to rely on researchers and expert judgements, accumulated and analyzed by an official standards organization. As described above, in this scenario there are various advantages to employing the expert-based approach, including the ability to cope with zero-day attacks, generalizability, scalability and objectivity.

\section{Conclusion and Future Work}\label{sec:conclusion_and_future_work}

Based on device complexity and the associated traffic predictability, this study presents the first step toward explaining and quantifying the detectability of IoT-related attacks. In accordance with the ideas presented in~\cite{surowiecki2005wisdom} and in order to reduce the overhead associated with the data-driven approach, we opted for an expert-based approach. Our proposed method leverages the collective wisdom and experience of cyber-security researchers and practitioners, and can serve as an interim solution to \emph{cold start} problems, in which preliminary attack detectability assessments are required, yet a large-scale high-quality correctly labeled dataset is not available.

In this study, we \emph{(a)} quantitatively explored the predictability of IoT network traffic, which was taken for granted thus far, \emph{(b)} designed a detailed taxonomy for IoT attack detection, \emph{(c)} proposed a means of quantifying the values of the taxonomy's features for a given IoT model, \emph{(d)} derived four detectability models which correspond to common attack scenarios, \emph{(e)} evaluated our method for \emph{DDoS flooding} attacks using commercial IoT devices, and \emph{(f)} compared our results with past research. 

Regarding network traffic predictability, we found that \emph{(a)} the traffic of IoT devices is significantly more predictable than of non-IoT devices, \emph{(b)} variability in traffic predictability does exist among disparate IoT models, and \emph{(c)} the IoT traffic predictability is correlated with static device characteristics (e.g., the number of sensors). Concerning our non-trivial selection of an expert-based approach (rather than a data-driven one) to quantify in advance the detectability of IoT-related attacks, we found that \emph{(d)} despite differing backgrounds and experience levels, the agreement among the cyber-security experts was relatively high. In addition, \emph{(e)} the weights which were assigned to the sub-categories and features using our expert-based method align with the importance of corresponding features from past data-driven studies on IoT attack detection. Additional findings are that \emph{f)} the D-Scores which were attained using our expert-based method are aligned with data-driven attack detection performance from past research, and \emph{(g)} higher detectability is associated with IoT models whose baseline traffic is easier to profile, i.e., they communicate less frequently, with less variability in packet sizes and with fewer destination ports. 

In future research we will take additional steps toward further explaining and quantifying the detectability of IoT-related attacks, while focusing on the following challenges:

\begin{itemize}[leftmargin=*]
\item Broaden the scope of quantitatively evaluating our method by means of, e.g., additional attack scenarios, more questionnaire responses to be collected from further security experts, and additional data-driven methods and experimental detection results to compare with.
\item Analyze the ability of malicious actors to execute adversarial attacks. That is, an attacker who is familiar with the D-Score calculation method and parameters ($\alpha, \beta, \delta$, etc.) might use this knowledge to optimize an attack by, e.g., maximizing the frequency of outbound packets as part of a DDoS attack, while remaining ``under the radar”.
\item If the calculated D-Score of a certain IoT model to an attack scenario of interest is relatively low, yet the organization decides to deploy it, an option we consider to explore in future research is to leverage an IoT model's D-Score and take its value into account while setting the anomaly detection threshold of \emph{this} IoT model. That is, a deployed IoT model with a low D-Score is advised to use a low anomaly threshold, such that attacks would not be missed (i.e., to increase the TPR), while a different IoT model with a high D-Score can be associated with a relatively high threshold (to decrease the FPR).
\end{itemize}

\section*{Acknowledgment}

The authors thank the kind cyber-security experts who completed our questionnaire, thus enabled this research.

\ifCLASSOPTIONcaptionsoff
  \newpage
\fi

\bibliographystyle{IEEEtran}
\bibliography{d_score_bib.bib}

\newpage

\begin{IEEEbiography}[{\includegraphics[width=1in,height=1.25in,clip,keepaspectratio]{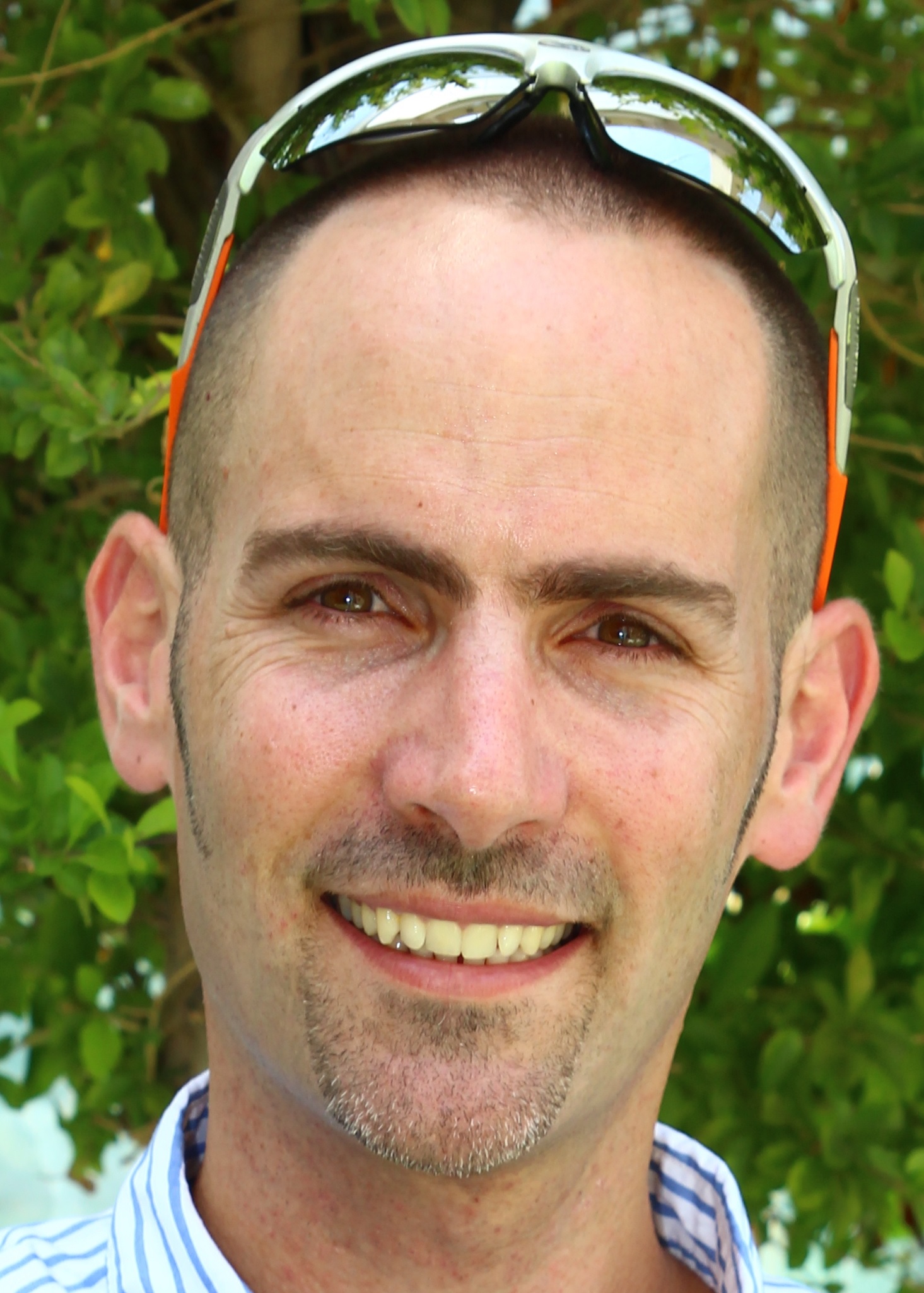}}]{Yair Meidan} is a PhD candidate in the Department of Software and Information Systems Engineering (SISE) at Ben-Gurion University of the Negev (BGU). His research interests include machine learning and IoT security. Contact him at yairme@post.bgu.ac.il.
\end{IEEEbiography}

\begin{IEEEbiography}[{\includegraphics[width=1in,height=1.25in,clip,keepaspectratio]{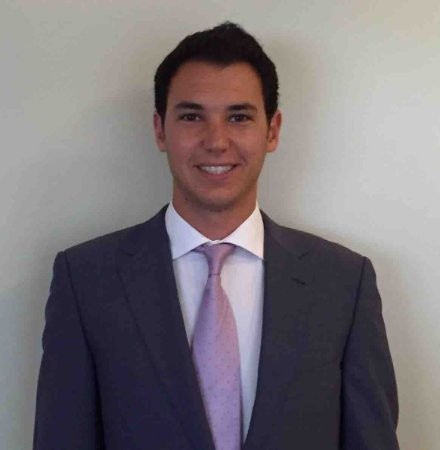}}]{Daniel Benatar} is an MSc student in the SISE Department at BGU. His research interests include machine learning and cyber-security. Contact him at benatar@post.bgu.ac.il.
\end{IEEEbiography}

\begin{IEEEbiography}[{\includegraphics[width=1in,height=1.25in,clip,keepaspectratio]{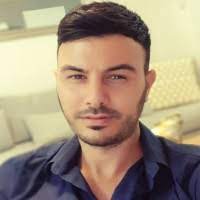}}]{Ron Bitton} is a PhD candidate in the SISE Department at BGU. His research interests include machine learning, cyber-security, and cyber-risk management. Contact him at ronbit@post.bgu.ac.il.
\end{IEEEbiography}

\begin{IEEEbiography}[{\includegraphics[width=1in,height=1.25in,clip,keepaspectratio]{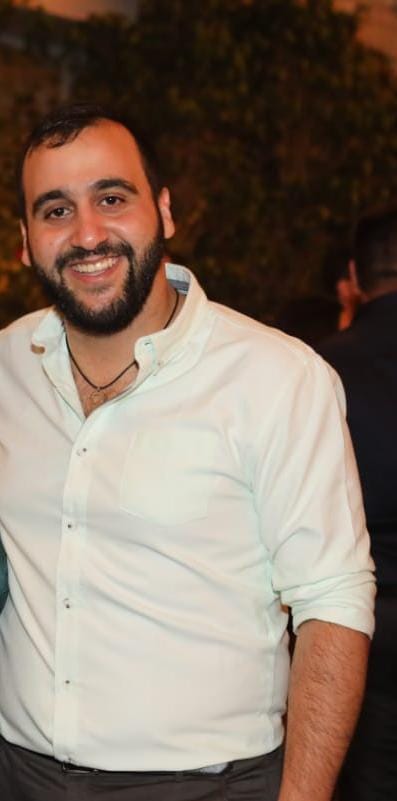}}]{Dan Avraham} is an MSc student in the SISE Department at BGU. His research interests include machine learning and cyber-security. Contact him at danavra@post.bgu.ac.il.
\end{IEEEbiography}

\begin{IEEEbiography}[{\includegraphics[width=1in,height=1.25in,clip,keepaspectratio]{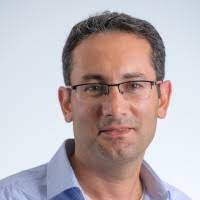}}]{Asaf Shabtai} is a professor in the SISE Department at BGU. His research interests include computer and network security, and machine learning. Shabtai received a PhD in information systems from BGU. Contact him at shabtaia@bgu.ac.il.
\end{IEEEbiography}

\newpage
\appendices

\section{Analysis of Traffic Predictability using an Additional Dataset (UNSW)~\cite{Sivanathan2019classifying}}\label{apndx:UNSW}
\begin{figure*}[!htbp]
\begin{minipage}{.45\linewidth}
\centering
\subfloat[]{\label{subfig_apndx:IoT_vs_Non_IoT_1}\includegraphics[height=0.2\textheight, trim={0cm 0cm 0cm 0cm},clip]
{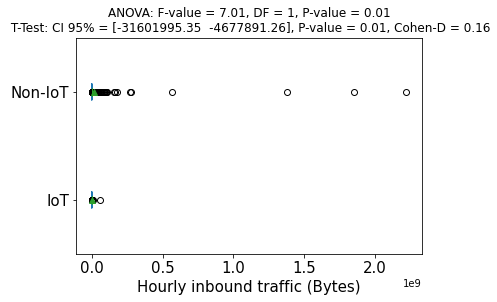}}
\end{minipage}
\hspace{\fill}
\begin{minipage}{.45\linewidth}
\centering
\subfloat[]{\label{subfig_apndx:IoT_vs_Non_IoT_2}\includegraphics[height=0.2\textheight, trim={0cm 0cm 0cm 0cm},clip]
{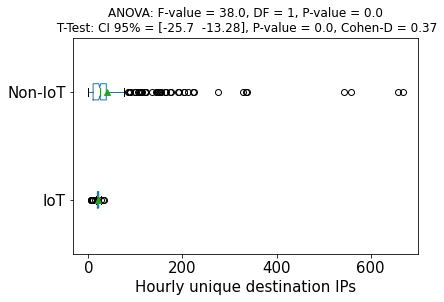}}
\end{minipage}
\vskip\baselineskip
\begin{minipage}{.45\linewidth}
\centering
\subfloat[]{\label{subfig_apndx:among_IoT_1}\includegraphics[height=0.2\textheight, trim={0cm 0cm 0cm 0cm},clip]
{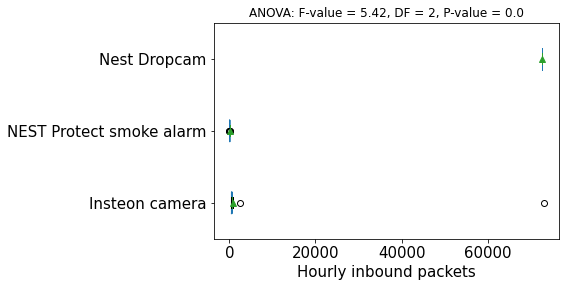}}
\end{minipage}
\hspace{\fill}
\begin{minipage}{.45\linewidth}
\centering
\subfloat[]{\label{subfig_apndx:among_IoT_2}\includegraphics[height=0.2\textheight, trim={0cm 0cm 0cm 0cm},clip]
{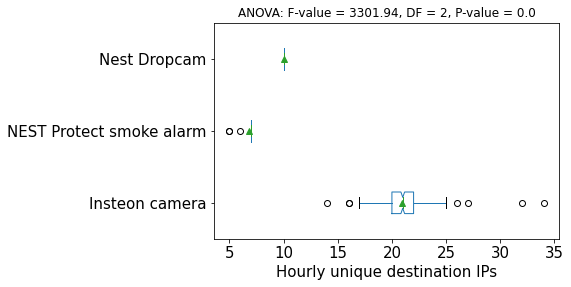}}
\end{minipage}
\vskip\baselineskip
\begin{minipage}{.45\linewidth}
\centering
\subfloat[]{\label{subfig_apndx:IoT_corr_1}\includegraphics[height=0.2\textheight, trim={0cm 0cm 0cm 0cm},clip]
{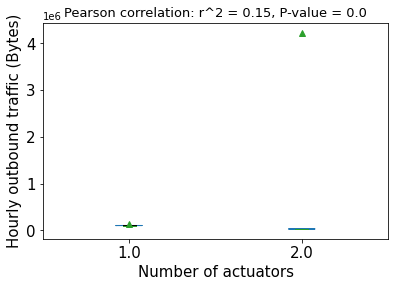}}
\end{minipage}
\hspace{\fill}
\begin{minipage}{.45\linewidth}
\centering
\subfloat[]{\label{subfig_apndx:IoT_corr_2}\includegraphics[height=0.2\textheight, trim={0cm 0cm 0cm 0cm},clip]
{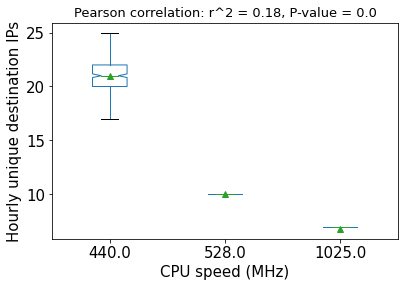}}
\end{minipage}
\caption{Traffic predictability analysis using UNSW data~\cite{Sivanathan2019classifying}: Comparison between IoT and non-IoT devices (Fig.~\ref{subfig_apndx:IoT_vs_Non_IoT_1} and~\ref{subfig_apndx:IoT_vs_Non_IoT_2}), comparison among IoT models (Fig.~\ref{subfig_apndx:among_IoT_1} and~\ref{subfig_apndx:among_IoT_2}), and correlation with IoT model complexity (Fig.~\ref{subfig_apndx:IoT_corr_1} and~\ref{subfig_apndx:IoT_corr_2}). In this dataset, the IoT devices comprised of Nest Dropcam, Nest Protect smoke alarm, and Insteon camera, while the non-IoT devices comprised of Samsung Galaxy Tab, iPhone, a laptop, two Macbooks, and two Android phones.}
\label{fig_apndx:UNSW}
\end{figure*}

\end{document}